\documentclass[11pt]{article}
\usepackage[utf8]{inputenc}
\usepackage{physics}
\usepackage{mathrsfs}
\usepackage[colorlinks=true,citecolor=blue,linkcolor=blue]{hyperref}
\usepackage[english]{babel}
\usepackage[T1]{fontenc}
\usepackage[left=2cm,right=2cm,top=2cm,bottom=2cm]{geometry}
\usepackage{amsmath,amssymb,amsthm,dsfont}
\usepackage{amsfonts}
\usepackage{graphicx}
\usepackage{subfig}
\usepackage{wrapfig}
\usepackage{slashed}
\usepackage{babel}
\begin{document}
	\title{\vspace{-3cm}
		\hfill\parbox{4cm}{\normalsize \textit{}}\\
		\vspace{1cm}
		{Laser-assisted scattering of a muon neutrino by an electron within the Electroweak theory}}
	\vspace{2cm}
	\author{S. El Asri$^1$, S. Mouslih$^{1,2}$, M. Ouali$^1$, S. Taj$^1$, B Manaut$^{1,}$ \\
		{\it {\small$^1$ Sultan Moulay Slimane University, Polydisciplinary Faculty,}}\\
		{\it {\small Laboratory of Research in Physics $\&$ Engineering Sciences, Team of Modern and Applied Physics,}}\\
		{\it {\small Beni Mellal, 23000, Morocco.}}
		\\			
	{\it {\small$^2$Faculty of Sciences and Techniques, 
		Laboratory of Materials Physics (LMP),
		Beni Mellal, 23000, Morocco.}}		
	}
	\maketitle \setcounter{page}{1}




\begin{abstract}
In accordance with the electroweak theory, we perform, in the first-Born approximation, a detailed analytical treatment of the differential cross-section (DCS) for the elastic scattering process $e^{-}+\nu_{\mu}\longrightarrow e^{-}+\nu_{\mu}$ in both absence and presence of a circularly polarized laser field. This scattering process is examined by using Dirac-Volkov wave functions for charged particles within an external laser field. The theoretical results obtained for the differential cross-section without laser field are compared with other theoretical results within the framework of Fermi theory. These results revealed to us the most important points that contribute to the understanding of the electroweak theory role in solving the problem of unitarity violation of the differential cross-section in Fermi theory. The differential cross section is modified significantly by inserting the electrons into an electromagnetic field. In addition, the influence of the laser strength and its frequency on the photons exchange between the colliding system and the laser field are also included and discussed. These effects are heavily related to the order and argument of the ordinary Bessel functions introduced in the theoretical calculation. 
\end{abstract}

\maketitle
\section{Introduction}
Scientists have been interested in the study of intense and effective interactions of charged particles with the electromagnetic field since its discovery in 1960s until today. This interest has contributed to a more precise understanding of the atomic structure of matter. Laser technology has achieved a great development in recent years, particularly in increasing its intensity to $10^{22}~\text{w/cm$^{-2}$}$\cite{highlaser} and shortening its pulse duration \cite{laser,laser1,laser3}. In addition, lasers are indispensable tools for the study of physical processes in various fields such as atomic, particle and nuclear physics. Then, it has become the center of attention of scientists and researchers on the study of scattering processes in the presence of the laser field. Due to the availability of high-intensity lasers, it is possible to investigate experimentally these scattering processes and to observe multi-photon processes \cite{test1,test2,test3}. In order to have an overview of the laser-assisted scattering processes, we would like to refer the reader to some books \cite{faisal,mittleman,fedorov} which deal with theses type of processes. At the beginning, the researchers were enthusiastic, and they concentrated on the study of laser-assisted scattering processes in both atomic physics and quantum electrodynamics.
For instance, in atomic physics, enormous works studied the diffusion of charged particles by an atom with the help of an electromagnetic field. In refs \cite{szymanowski1,szymanowski2,manaut2003}, the authors studied Mott scattering by taking into account the spin and relativistic effect in the presence of a circularly polarized laser field. The same process was treated in Refs \cite{li2003,attaourti2004} in the presence of a linearly and elliptically polarized laser field in the lowest order. In addition to atomic physics processes, various authors studied laser-assisted scattering processes in quantum field theory (QED). The relativistic elastic scattering of an electron by a proton has been analyzed in the presence of a linearly and circularly polarized laser field in \cite{dahiri,liu2014,wang2019}. In addition, in Refs \cite{du2018,Nonresonant,Yahya}, the authors have studied new phenomena of scattering of an electron by a muon in the laser field with different polarizations. 

In recent years, the treatment of scattering processes in electroweak theory, which is discovered in 1967 by Sheldon Lee Glashow, Abdus Salam and Steven Weinberg \cite{Weinberg,Salam}, has become interesting to many researchers and physicists. This theory is based on unifying the QED and the weak interactions in the standard model by using the gauge group $\text{SU(2)}\otimes \text{U(1)}$, and it is also capable of predicting the masses of gauge bosons $Z^{0}$ and $W^{\mp}$.  Thus, it has played an important role not only in providing very interesting physical information that has paved the way for new works in the community of particle physics but also in understanding the interaction between matter and lasers. Moreover, numerous authors have studied some scattering and decay processes in the electroweak theory within a laser field \cite{Muller,jakha,hadwdecay,ouhammou,Ouali}. 

In this work, we will study one of the most important processes in physical research, namely the process of elastic scattering of the muon neutrino by an electron ($e^{-}+\nu_{\mu}\longrightarrow e^{-}+\nu_{\mu}$) under the influence of an external field. This process greatly contributed to the discovery and the understanding of the electroweak theory \cite{vilain} as it provided the first experimental evidence for the existence of weak neutral current interactions (see Refs\cite{hasert,tomalak,marciano} for more details). Ten years ago, the same process was studied under the influence of a linearly polarized electromagnetic field in the context of the Fermi theory \cite{bai2012}. The authors of this work found that the distributions of a multi-photon energy transfer spectrum and the Differential Cross-Section are significantly changed when a strong laser field is present. Recently, we have also studied this process in the framework of Fermi theory, but in the presence of a circularly polarized laser field \cite{ElAsri}. Our results show that the laser field has a significant effect on the DCS as well as on the process of photons exchange between the laser and scattering process. However, this treatment cannot be convenient in all situations, in particular at high energies, and this is due to the known limitations of the Fermi theory. The aim of this paper is to study the process $e^{-}+\nu_{\mu}\longrightarrow e^{-}+\nu_{\mu}$ in the framework of electroweak theory by introducing the $Z$-boson propagator, and to solve the problem of unitarity violation of the cross-section which arises in the Fermi theory at high energies. In addition, our main contribution is that we have presented for the first time a detailed calculation of this process in the absence and presence of the laser field. We have also compared the DSCs in the two theories (Fermi and electroweak theory), and we have examined the effect of the laser field parameters on the DCS. 

The rest of this paper is organized as follows: Firstly, in section \ref{sec:free}, we describe the methodology used to evaluate the S-matrix element in the first Born approximation as well as the expression of the DCS in the absence of an external laser field. In section \ref{sec:Laser-assisted}, we give detailed theoretical calculations related to DCS under the influence of an external electromagnetic field. In section \ref{NUMERICAL RESULTS}, we present the numerical results of the DCS in both absence and presence of the external electromagnetic field. Finally, section \ref{Conclusion} is devoted to a brief summary and conclusion. We note that throughout this paper, we have used the natural units $c=\hbar=1$ where $c$ is the velocity of light in vacuum, and the Minkowski metric tensor  $g^{\mu\nu}=\text{diag}(1,-1,-1,-1)$.
\section{Laser-free scattering process}\label{sec:free}
\begin{figure}[hptb]
 		\centering
 		\includegraphics[scale=0.5]{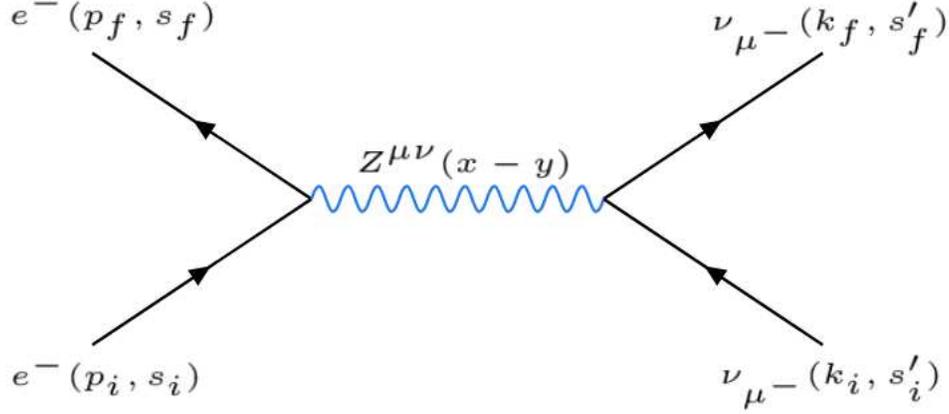}
 		\caption{Tree-level t-channel Feynman diagram of the elastic scattering of a muon neutrino by an electron in the framework of electroweak theory. Time is oriented from the bottom to the top.}\label{diagram}
 	\end{figure}

In this paper, we consider the elastic scattering of a muon neutrino by an electron in the framework of electroweak theory. This process can be structured as follows:
 	\begin{equation}\label{process}
 	e^{-}(p_{i})+\nu_{\mu}(k_{i})\longrightarrow e^{-}(p_{f})+\nu_{\mu}(k_{f}),
 	\end{equation}
where the arguments $p_{i}$ and $p_{f}$ are the four-vector momentum of the incoming and outgoing electrons. $k_{i}$ and $k_{f}$ are successively the free four-vector momentum of the incident and scattered muon neutrino. We assume that this process is an electroweak interaction process as it is mediated by a neutral weak gauge $Z$-boson. It can be described by the lowest order Feynman diagram represented in figure \ref{diagram}.
The overall coupling between the $Z$-boson and fermions are defined as follows:
 	\begin{equation}
 	\text{vertex}~Z\text{-}e\text{-}e=\dfrac{-ig}{2\cos(\theta_{W})}\gamma_{\nu}(g_{V}-g_{A}\gamma_{5});~~\text{vertex}~Z\text{-}\nu_{\mu}\text{-}\nu_{\mu}=\dfrac{ig}{4\cos(\theta_{W})}\gamma_{\mu}(1-\gamma^{5}).
 	\end{equation}
By using the Feynman rules, the $S$-matrix element for the studied scattering process in the first order can be expressed as follows: 
 	\begin{equation}\label{S-matrix}
 	S_{fi}=\dfrac{g^{2}}{8\cos^{2}(\theta_{W})}\int\int d^{4}x~d^{4}y \Big[\overline{\psi}^{\textsl{f}}_{e^{-}}(x)\gamma_{\nu}(g_{V}-g_{A}\gamma_{5})\psi^{\textsl{i}}_{e^{-}}(x)\Big]Z^{\mu\nu}(x-y)\Big[\overline{\psi}^{\textsl{f}}_{\nu_{\mu}}(y)\gamma_{\mu}(1-\gamma_{5})\psi^{\textsl{i}}_{\nu_{\mu}}(y)\Big],
 	\end{equation}
where $g$ is the electroweak coupling constant, and  $\theta_{W}$ is the Weinberg angle. $g_{V}=1/2-2\sin^{2}(\theta_{W})$ and $g_{A}=-1/2$ are the vector and axial-vector coupling constants, respectively.
We have used the corrected values (at the loop level) for $g_{V}$ and $g_{A}$ such that: $g_{V}=0.043 \mp 0.063$ and $g_{A}=-0.545 \mp0.056$  \cite{greiner}.
The incoming and the outgoing muon neutrinos are described by Dirac wave functions normalized to the volume V, and they are expressed as follows:
 	\begin{equation}\label{wave function of neutrino}
 	\begin{split}
 	&\psi^{\textsl{i}}_{\nu_{\mu}}(x)=\dfrac{1}{\sqrt{2E_{i}V}}u_{\nu_{\mu}}(k_{i},s'_{i})e^{-ik_{i}.x},\\
 	&\psi^{\textsl{f}}_{\nu_{\mu}}(x)=\dfrac{1}{\sqrt{2E_{f}V}}u_{\nu_{\mu}}(k_{f},s'_{f})e^{-ik_{f}.x},
 	\end{split}
 	\end{equation}
 	where $E_{i}$ and $E_{f}$ are successively the total energies of the incident and scattered muon neutrino.
 	 $u_{\nu_{\mu}}(k_{i,f},s'_{i,f})$ denotes the bispinor of the free muon neutrino. The four-momentum $k_{i,f}$ and the spin $s'_{i,f}$ satisfy the following equation:
 	\begin{equation}
 	\sum_{s'_{i,f}}\bar{u}_{\nu_{\mu}}(k_{i,f},s'_{i,f})u_{\nu_{\mu}}(k_{i,f},s'_{i,f})= \slashed{k}_{i,f},
 	\end{equation}
where the two subscripts i and f correspond to the initial and final states, respectively. The electron in the initial and final states is described by the Dirac-free states as follows:
\begin{equation}\label{wave function of electron}
 	\begin{split}
 	&\psi^{\textsl{i}}_{e^{-}}(x)=\dfrac{1}{\sqrt{2p_{i}^{0}V}}u_{e^{-}}(p_{i},s_{i})e^{-ip_{i}.x},\\
 	&\psi^{\textsl{f}}_{e^{-}}(x)=\dfrac{1}{\sqrt{2p_{f}^{0}V}}u_{e^{-}}(p_{f},s_{f})e^{-ip_{f}.x},
 	\end{split}
 	\end{equation}
  	where $u_{e^{-}}(p_{i,f},s_{i,f})$ are the Dirac bispinors of the incident and scattered electron outside the laser field. The free four-momentum of the electron, $p_{i,f}$, and its spins $s_{i,f}$ satisfy the following condition  $\sum_{s_{i,f}}\bar{u}_{e^{-}}(p_{i,f},s_{i,f})u_{e^{-}}(p_{i,f},s_{i,f})= \slashed{p}_{i,f}+m $,
where $m$ is the rest mass of the free electron. Further, $p_{i}^{0}$ and $p_{f}^{0}$ are, respectively, the total energies of the incoming and outgoing electrons. $Z^{\mu\nu}(x-y)$ is the $Z$-boson Feynman propagator \cite{greiner}, which is defined as follows:
\begin{equation}
 	Z^{\mu\nu}(x-y)=\int\dfrac{d^{4}q}{(2\pi)^{4}}\dfrac{e^{-iq(x-y)}}{q^{2}-M_{Z}^{2}}\Big[-ig^{\mu\nu}+i(1-\zeta)\dfrac{q^{\mu}q^{\nu}}{M_{Z}^{2}}\Big].
 	\end{equation}
Since all the observables, such as the DCS, do not depend on the gauge choice \cite{MohaPhLettB}, we will choose the Feynman gauge, which corresponds to $ \zeta = 1$. Therefore, the Feynman propagator for the $Z$-boson becomes:	
 \begin{equation}\label{feynman propagator}
 	Z^{\mu\nu}(x-y)=-i\int\dfrac{d^{4}q}{(2\pi)^{4}}\dfrac{g^{\mu\nu}}{q^{2}-M_{Z}^{2}}e^{-iq(x-y)}.
 	\end{equation}
To obtain the analytical results, we proceed as follows: We insert equations (\ref{wave function of neutrino}), (\ref{wave function of electron}) and (\ref{feynman propagator}) into the formula of the $S$-matrix (\ref{S-matrix}). After some manipulations, the $S$-matrix becomes as follows:
 	\begin{equation}\label{S-matrix1}
 	S_{fi}=\dfrac{-ig^{2}}{8cos^{2}(\theta_{W})}\dfrac{1}{\sqrt{16E_{i}E_{f}p_{i}^{0}p_{f}^{0} V^{4}}}\int\int\int d^{4}xd^{4}y\dfrac{d^{4}q}{(2\pi)^{4}}\times\frac{\mathcal{M}_{fi}}{q^{2}-M_{Z}^{2}}~ e^{i(p_{f}-p_{i}-q).x}~e^{i(k_{f}-k_{i}+q).y},
 	\end{equation}
 	where
 \begin{equation}
 M_{fi}=[\bar{u}_{e^{-}}(p_{f},s_{f})\gamma^{\mu}(g_{V}-g_{A}\gamma^{5})u_{e^{-}}(p_{i},s_{i}) \big] \big[\bar{u}_{\nu_{\mu}}(k_{f},s'_{f})\gamma_{\mu}(1-\gamma_{5})u_{\nu_{\mu}}(k_{i},s'_{i}) \big].
 \end{equation}
The integration over $d^{4}x$, $d^{4}y$ and $d^{4}q$ in eq. (\ref{S-matrix1}) can be calculated by using standard techniques \cite{greiner}. Therefore, the scattering matrix element becomes:
 \begin{equation} 
 S_{fi}=\dfrac{-ig^{2}}{8cos^{2}(\theta_{W})}\dfrac{1}{\sqrt{16E_{i}E_{f}p_{i}^{0}p_{f}^{0} V^{4}}} \frac{(2\pi)^{4}\delta^{4}(p_{f}-p_{i}+k_{f}-k_{1})}{q^{2}-M_{Z}^{2}}\mathcal{M}_{fi},
 \end{equation}
where $q=p_{f}-p_{i}$ is the relativistic momentum transfer in the absence of the laser field. In the case of unpolarised DCS, we have to sum over the final spin states and average over the initial ones. Further, The neutrinos exist in a single state of negative helicity \cite{greiner}. Thus, To express the non-polarised DCS without laser,  we multiply the squared $S$-matrix element, $|S_{fi}|^2$, by the density of final states, and divide it by the flux of the incident particles $|J_{\text{inc}}|$. Thus, the unpolarized DCS is expressed as follows:
\begin{center}
\begin{figure}[hbtp]
\centering
\includegraphics[scale=0.63]{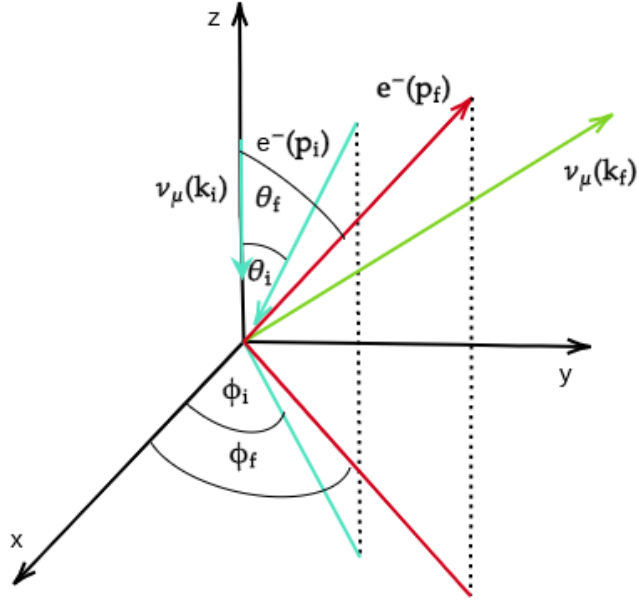}
\caption{Geometry of the collision process $e^{-}+\nu_{\mu}\longrightarrow e^{-}+\nu_{\mu}$. The incoming muon neutrino $\nu_{\mu}(k_{i})$ moves along the opposite direction of the $z$-axis.  The incident electron moves in a general geometry with spherical coordinates $\theta_{i}$ and $\phi_{i}$. The outgoing muon neutrino and electron move in $xyz$ space.}
\label{fig1}
\end{figure}
\end{center}
\begin{equation}
d\overline{\sigma}(e^{-} \nu_{\mu} \rightarrow e^{-} \nu_{\mu})=~V\int\dfrac{d^{3}p_{f}}{(2\pi)^{3}}~V\int\dfrac{d^{3}k_{f}}{(2\pi)^{3}}~\dfrac{1}{2}\sum_{s'_{i,f},s_{i,f}}\dfrac{|S_{fi}(e^{-} \nu_{\mu} \rightarrow e^{-} \nu_{\mu})|^{2}}{ T|J_{inc}|},
\end{equation}
where $|J_{inc}|=\sqrt{(k_{i}.p_{i})^{2}-m_{\nu_{\mu}}^{2}m^{2}}/(E_{i} p_{i}^{0} V )$ denotes the incoming particle current in the laboratory system. We use $g^{2} = 8\textit{G}_{F} M^{2}_{Z} \cos^{2}(\theta_{W})/ \sqrt{2}$, with $\textit{G}_{F}=(1.166~37\pm0.000~02)\times10^{-11}~\text{MeV}^{-2}$ is the Fermi coupling constant measured from the muon decay \cite{pdg2020} and by using the following relations: $d^{3}p_{f}=|\textbf{p}_{f}|^{2}d|\textbf{p}_{f}|d\Omega $ and $\delta^{4}(p_{f}+k_{f}-p_{i}-k_{i})=\delta^{0}(p_{f}^{0}+E_{f}-p_{i}^{0}-E_{i})\delta^{3}(\textbf{p}_{f}+\textbf{k}_{f}-\textbf{p}_{i}-\textbf{k}_{i})$. After simplifications, we get:
\begin{equation}\label{DCSdiverges}
\dfrac{d\overline{\sigma}}{d\Omega}=\dfrac{\textit{G}_{F}^{2}}{64(2\pi)^{2}(k_{i}.p_{i})}\frac{M_{Z}^{4}}{(q^{2}-M_{Z}^{2})^{2}}\int\dfrac{|\textbf{p}_{f}|^{2}d|\textbf{p}_{f}|}{E_{f}p_{f}^{0}} \delta^{0}(p_{f}^{0}+E_{f}-p_{i}^{0}-E_{i})\sum_{s'_{i,f},s_{i,f}}|M_{fi}|^{2}\bigg|_{\textbf{p}_{f}+\textbf{k}_{f}-\textbf{p}_{i}-\textbf{k}_{i}=0.}
\end{equation}
The remaining integral over $ d|\textbf{p}_{f}|$ can be solved with the help of the formula below \cite{greiner}:
 	\begin{align}\label{familiarformula}
 	\int dyf(y)\delta(g(y))=\dfrac{f(y)}{|g'(y)|}\bigg|_{g(y)=0.}
 	\end{align}
 	The $Z$-boson propagator given by equation (\ref{feynman propagator}) does not take into account the fact that the $Z$-boson is an unstable particle, and for this reason the DCS diverges at $q^{2}=M_{Z}^{2} $ (equation (\ref{DCSdiverges})). To avoid this divergence, We will use the following transformation \cite{greiner}:
 	\begin{equation}
 	\frac{1}{q^{2}-M_{Z}^{2}}\rightarrow 	\frac{1}{q^{2}-M_{Z}^{2}+i~M_{Z}~\Gamma_{Z}},
 	\end{equation}
 where $\Gamma_{Z}=(2.4952\pm 0.0023)~\text{GeV}$ is the total decay rate of $Z$-boson \cite{PDGroup2020}. Thus, we get:
 	\begin{equation}\label{freedcs}
 	\begin{split}
 	\dfrac{d\overline{\sigma}}{d\Omega}(e^{-} \nu_{\mu} \rightarrow e^{-} \nu_{\mu})=\dfrac{\textit{G}_{F}^{2}}{ 256 \pi^{2} p_{f}^{0}E_{f}}\dfrac{M_{Z}^{4}}{\big(q^{2}-M_{Z}^{2}\big)^{2}+M_{Z}^2~\Gamma_{Z}^2}\dfrac{|\textbf{p}_{f}|^{2}}{(k_{i}.p_{i}) |g'(|\textbf{p}_{f}|)|}\sum_{s'_{i,f},s_{i,f}}|M_{fi}|^{2},
 	\end{split}
 	\end{equation}
 	where
 	\begin{equation}
 	g'(|\textbf{p}_{f}|)=\dfrac{|\textbf{p}_{f}|}{\sqrt{|\textbf{p}_{f}|^{2}+m^{2}}}+\dfrac{|\textbf{p}_{f}|+|\textbf{k}_{i}|\cos(\theta_{f})-|\textbf{p}_{i}|F(\theta_{i},\theta_{f},\phi_{i},\phi_{f})}{E_{f}},
 	\end{equation}
 	and 
 	\begin{equation}
 	\begin{split}
 	F(\theta_{i},\theta_{f},\phi_{i},\phi_{f})=&\cos(\phi_{i})\sin(\theta_{i})\cos(\phi_{f})\sin(\theta_{f})+\sin(\theta_{i})\sin(\phi_{i})\sin(\theta_{f})\sin(\phi_{f})\\
 	&+\cos(\theta_{i})\cos(\theta_{f}).
 	\end{split}
 	\end{equation}
\section{Laser-assisted scattering process}\label{sec:Laser-assisted}
In this section, we study the process (\ref{process}) in the presence of a laser field. In this situation, we consider that the electron is immersed in a circularly polarized monochromatic laser field. The latter is described by its classical four-potential $A^{\mu}(\phi)$ which verifies the Lorentz gauge condition $\partial A^{\mu}(\phi)=0$. So, the classical four-potential can be expressed in a unified notation such as:
\begin{equation}
A^{\mu}(x)=|\textbf{a}| \big[\eta_{1}^{\mu} \cos(k.x)+ \eta_{2}^{\mu} \sin(k.x)\big],
\end{equation}
where $ k=(\omega,\textbf{k})$ is the wave four-vector, and $\omega$ its frequency.  $|\textbf{a}|=\xi_{0}/\omega$  is the magnitude of the four-potential with $ \mathcal{E}_{0} $ is the electric field strength. The polarization four-vectors $\eta_{1}^{\mu} $ and $ \eta_{2}^{\mu}$ are orthogonal and equal in magnitude, and they are given by:
\begin{equation}
\eta_{1}^{\mu}=(0,1,0,0),~~~\eta_{2}^{\mu}=(0,0,1,0).
\end{equation}
These quantities meet the following conditions: $\eta_{1}^{2}=\eta_{2}^{2}=-1$ and $(\eta_{1}.\eta_{2})=0$. According to the Lorentz gauge condition $k_{\mu}.A^{\mu}=0$, we obtain $(k.\eta_{1})=0$ and  $(k.\eta_{2})=0$. Inside the laser field, the incoming and outgoing electrons are described by Dirac-Volkov states \cite{volkov} as follows:
\begin{equation}\label{Sqfi}
\begin{split}
&\psi^{\textsl{i}}_{e^{-}}(x)=\bigg[1+\dfrac{e\slashed{k}\slashed{A}}{2(k.p_{i})}\bigg]\frac{u(p_{i},s_{i})}{\sqrt{2Q_{i}V}}\times e^{iS(q_{i},x)},\\
&\psi^{\textsl{f}}_{e^{-}}(x)=\bigg[1+\dfrac{e\slashed{k}\slashed{A}}{2(k.p_{f})}\bigg]\frac{u(p_{f},s_{f})}{\sqrt{2Q_{f}V}}\times e^{iS(q_{f},x)},
\end{split}
\end{equation}
where $Q_{i} $ and $Q_{f}$ are the total energies of the incident and scattered electron in the presence of the electromagnetic field.
In equation (\ref{Sqfi}), the arguments of the exponential terms are expressed by:
\begin{equation}
\begin{split}
& S(q_{i},x)=-q_{i}.x-\dfrac{e|\textbf{a}|(\eta_{1}.p_{i})}{(k.p_{i})}\sin(k.x)+\dfrac{e|\textbf{a}|(\eta_{2}.p_{i})}{(k.p_{i})}\cos(k.x),\\
& S(q_{f},x)=-q_{f}.x-\dfrac{e|\textbf{a}|(\eta_{1}.p_{f})}{(k.p_{f})}\sin(k.x)+\dfrac{e|\textbf{a}|(\eta_{2}.p_{f})}{(k.p_{f})}\cos(k.x),
\end{split}
\end{equation}
where $ q_{i,f}$ is the effective momentum which is related to its corresponding four-momentum outside the laser field by the following relation:
\begin{equation}
q_{i,f}=p_{i,f}+\frac{e^{2}|\textbf{a}|^{2}}{2(k.p_{i,f})}k.
\end{equation}
The square of this four-vector shows that the mass of the dressed electron depends on the laser field strength and its frequency. This dependence is illustrated in the following equation:
\begin{equation}
m_{*}^{2}=m^{2}+e^{2}|\textbf{a}|^{2},
\end{equation}
where, the quantity $m_{*}$ acts as an effective mass of the electron within the electromagnetic field. Due to its neutral electric charge, the muon neutrino will not interact with the electromagnetic field.
Therefore, it will be described by Dirac free state as given by equation (\ref{wave function of neutrino}). After some manipulations,  we find  the laser-assisted scattering matrix element for the elastic scattering of a muon neutrino by an electron can be expressed as follows:
 	\begin{equation}\label{Sfi}
 	\begin{split}
 	S_{fi}=&\dfrac{-i g^{2}}{8\cos^2(\theta_{W})~(q^{2}-M_{Z}^{2})}\dfrac{1}{\sqrt{16E_{i}E_{f}Q_{i}Q_{f} V^{4}}}\int d^{4}xd^{4}y\dfrac{d^{4}q}{(2\pi)^{4}} e^{i(q_{f}-q_{i}-q).x}~e^{i(k_{f}-k_{i}+q).y}~e^{-iz\sin(k.x-\varphi_{0})}\\
&\bigg[\bar{u}(p_{f},s_{f})\Big(1+\dfrac{e\slashed{A}\slashed{k}}{2(k.p_{f})}\Big)
 \gamma^{\mu}(g_{V}-g_{A}\gamma^{5})\Big(1+\dfrac{e\slashed{k}\slashed{A}}{2(k.p_{i})}\Big)u(p_{i},s_{i}) \bigg]\bigg[\bar{u}_{\mu_{\mu}}(k_{f},s'_{i})\gamma_{\mu}(1-\gamma_{5})u_{\nu_{\mu}}(k_{i},s'_{f}) \bigg]. 	
 	\end{split}
 	\end{equation}
In equation (\ref{Sfi}), we have used the following transformation:
 	\begin{equation}
 e^{i(S(q_{i},x)-S(q_{f},x))}=e^{i(q_{f}-q_{i}).x} ~ e^{-iz\sin(k.x-\varphi_{0})},
 	\end{equation}
 	  where
 	\begin{equation}\label{argument}
z=e |\textbf{a}|\sqrt{\bigg(\dfrac{\eta_{1}.p_{i}}{k.p_{i}}-\dfrac{\eta_{1}.p_{f}}{k.p_{f}}\bigg)^{2}+\bigg(\dfrac{\eta_{2}.p_{i}}{k.p_{i}}-\dfrac{\eta_{2}.p_{f}}{k.p_{f}}\bigg)^{2}},
\end{equation}
and
\begin{equation}
\varphi_{0}=\arctan\bigg[\frac{(\eta_{2}.p_{i})(k.p_{f})-(\eta_{2}.p_{f})(k.p_{i})}{(\eta_{1}.p_{i})(k.p_{f})-(\eta_{1}.p_{f})(k.p_{i})}\bigg].
\end{equation}
After some calculations, the expression of the scattering matrix element, $S_{fi}$, becomes as follows: 
 \begin{equation}
 \begin{split}
 S_{fi}=&\dfrac{-i g^{2}}{8\cos^2(\theta_{W})~(q^{2}-M_{Z}^{2})}\dfrac{1}{\sqrt{16E_{i}E_{f}Q_{i}Q_{f} V^{4}}}\ \int d^{4}xd^{4}y\dfrac{d^{4}q}{(2\pi)^{4}} e^{i(q_{f}-q_{i}-q)x}e^{i(k_{f}-k_{i}+q).y}~e^{-izsin(k.x-\varphi_{0})}\\
 &\times \Big[\bar{u}(p_{f},s_{f})\Big(\chi^{\mu}_{0}+\chi^{\mu}_{1}\cos(k.x)+\chi^{\mu}_{2}\sin(k.x)\Big) u(p_{i},s_{i}) \Big]\Big[\bar{u}_{\nu_{\mu}}(k_{f},s'_{f})\gamma_{\mu}(1-\gamma_{5})u_{\nu_{\mu}}(k_{i},s'_{i}) \Big],
 \end{split}
 \end{equation}
where
 	\begin{equation}
 	\begin{split}
 	&\chi^{\mu}_{0}=\gamma^{\mu}~(g_{V}-g_{A}\gamma^{5})+2~ C(p_{i})~ C(p_{f})~ |\textbf{a}|^{2}~k^{\mu}~\slashed{k}~(g_{V}-g_{A}\gamma^{5}),\\
 	& \chi^{\mu}_{1}=C(p_{i})|\textbf{a}| \gamma^{\mu}~(g_{V}-g_{A}\gamma^{5}) \slashed{k}~\slashed{\eta}_{1}+C(p_{f})|\textbf{a}|~\slashed{\eta}_{1}~\slashed{k}~\gamma^{\mu}~(g_{V}-g_{A}\gamma^{5}),\\
 	&\chi^{\mu}_{2}=C(p_{i})|\textbf{a}| \gamma^{\mu}~(g_{V}-g_{A}\gamma^{5}) \slashed{k}~\slashed{\eta}_{2}+C(p_{f})|\textbf{a}|~\slashed{\eta}_{2}~\slashed{k}~\gamma^{\mu}~(g_{V}-g_{A}\gamma^{5}).
 	\end{split}
 	\end{equation}
with $C(p_{i})=e/(2(k.p_{i}))$ and $C(p_{f})=e/(2(k.p_{f}))$. After integration over space-time and over $d^{4}q $, and by using the following transformation:
\begin{equation}
e^{iz\sin(\phi)}=\sum_{n=-\infty}^{n=+\infty} J_{n}(z) e^{in \phi},
\end{equation}
the scattering matrix element can be written in terms of ordinary Bessel function such that:
\begin{equation}\label{s-matrixlaser}
\begin{split}
S_{fi}=\dfrac{-i \textit{G}_{F}}{\sqrt{32E_{i}E_{f}Q_{i}Q_{f} V^{4}}} \dfrac{M_{Z}^{2}}{\big(q^{2}-M_{Z}^{2}\big)}\sum_{n=-\infty}^{+\infty}(2\pi)^{4}\delta^{4}(q_{f}+k_{f}-q_{i}-k_{i}-n k)~M^{n}_{fi},
\end{split}
\end{equation}
where $q=q_{f}-q_{i}-nk$ is the relativistic four-momentum transfer in the presence of the electromagnetic field, and  the integer $n$ indicates the number of laser photons that are emitted $(\text{if~n} > 0)$ or absorbed $(\text{if~ n} < 0)$ by the system. The quantity $M^{n}_{fi}$ can be expressed in terms of ordinary Bessel functions as follows:
 \begin{equation}\label{mfin}
M^{n}_{fi}=\Big[\bar{u}(p_{f},s_{f})\Big(\chi^{\mu}_{0} b_{n}(z)+\chi^{\mu}_{1} b_{1n}(z)+\chi^{\mu}_{2} b_{2n}(z)\Big)u(p_{i},s_{i}) \Big]\Big[\bar{u}_{\nu_{\mu}}(k_{f},s'_{f})\gamma_{\mu}(1-\gamma_{5})u_{\nu_{\mu}}(k_{i},s'_{i}) \Big],
\end{equation}
where
\begin{equation}
\begin{split}
   & b_{n}(z)=J_{n}(z)e^{in\varphi_{0}}\\\
   &b_{1n}(z)=\frac{1}{2}\big[ J_{n+1}(z)e^{i(n+1)\varphi_{0}}+J_{n-1}(z)e^{i(n-1)\varphi_{0}}\big]\\
 	&b_{2n}(z)=\frac{1}{2i}\big[ J_{n+1}(z)e^{i(n+1)\varphi_{0}}-J_{n-1}(z)e^{i(n-1)\varphi_{0}}\big],\\
\end{split}
\end{equation}
To calculate the DCS in the presence of an electromagnetic field, we follow the same procedure as in the absence of a laser field in the previous section. This yields:
  \begin{equation}
\dfrac{d\overline{\sigma}}{d\Omega}=\sum_{n=-\infty}^{+\infty}\dfrac{\textit{G}_{F}^{2}~M^4_{Z}}{64Q_{f}Q_{i}E_{f}E_{i}\big(q^{2}-M_{Z}^{2}\big)^{2}}\dfrac{|\textbf{q}_{f}|^{2}d|\textbf{q}_{f}|}{(2\pi)^{2}|J_{inc}|V} \delta^{0}(Q_{f}+E_{f}-Q_{i}-E_{i}-n\omega)\sum_{s'_{i,f},s_{i,f}}|M^{n}_{fi}|^{2}\bigg|_{\textbf{q}_{f}+\textbf{k}_{f}-\textbf{q}_{i}-\textbf{k}_{i}-n\textbf{k}=0.}
\end{equation}
Where  $|J_{inc}|=\sqrt{(k_{i}.q_{i})^{2}-m_{\nu_{\mu}}^{2}m^{2}}/(E_{i} Q_{i}^{0} V )$. We have used the formula (\ref{familiarformula}) to calculate the remaining integral over $d|\textbf{q}_{f}|$. Then, the final expression of the DCS becomes as follows:
\begin{equation}\label{dcswithlaser}
\Big(\dfrac{d\overline{\sigma}}{d\Omega}\Big)^{\text{with laser}}=\sum_{n=-\infty}^{+\infty}\dfrac{d\overline{\sigma}^{n}}{d\Omega},
\end{equation}
where the partial DCS can be written as:
\begin{equation}\label{IDCS}
\begin{split}
\dfrac{d\overline{\sigma}^{n}}{d\Omega}=&\frac{\textit{G}_{F}^{2}~M^4_{Z}}{64(2\pi)^{2}Q_{f}E_{f}\big(q^{2}-M_{Z}^{2}\big)^{2}+M_{Z}^2~\Gamma_{Z}^2}\dfrac{|\textbf{q}_{f}|^{2}}{(k_{i}.q_{i})|g'(|\textbf{q}_{f}|)|}\\
& \times \text{Tr}\bigg[(\slashed{p}_{f}+m)\Big(\chi^{\mu}_{0} b_{n}(z)+\chi^{\mu}_{1} b_{1n}(z)+\chi^{\mu}_{2} b_{2n}(z)\Big)(\slashed{p}_{i}+m)\Big(\overline{\chi}^{\mu}_{0} b^{*}_{n}(z)+\overline{\chi}^{\mu}_{1} b^{*}_{1n}(z)+\overline{\chi}^{\mu}_{2} b^{*}_{2n}(z)\Big)\bigg]\\
&\times\text{Tr}\bigg[(\slashed{k}_{f}+m_{\nu_{\mu}})\gamma_{\mu}(1-\gamma_{5})(\slashed{k}_{i}+m_{\nu_{\mu}})\gamma_{\nu}(1-\gamma_{5})\bigg],
\end{split}
\end{equation}
with 
 	\begin{equation}
 	\begin{split}
 	&\overline{\chi}^{\mu}_{0}=\gamma^{\nu}~(g_{V}-g_{A}\gamma^{5})+2~ C(p_{i})~ C(p_{f})~ |\textbf{a}|^{2}~k^{\nu}~\slashed{k}~(g_{V}-g_{A}\gamma^{5}),\\
 	& \overline{\chi}^{\mu}_{0}=C(p_{i}) |\textbf{a}|~\slashed{\eta}_{1}~\slashed{k}\gamma^{\nu}~(g_{V}-g_{A}\gamma^{5}) + C(p_{f}) |\textbf{a}|~\gamma^{\nu}~(g_{V}-g_{A}\gamma^{5})\slashed{k}~\slashed{\eta}_{1},\\
 	&\overline{\chi}^{\mu}_{0}=C(p_{i}) |\textbf{a}|~\slashed{\eta}_{2}~\slashed{k}\gamma^{\nu}~(g_{V}-g_{A}\gamma^{5}) +C(p_{f}) |\textbf{a}|~\gamma^{\nu}~(g_{V}-g_{A}\gamma^{5})\slashed{k}~\slashed{\eta}_{2}.
 	\end{split}
 	\end{equation}
 The computation of the traces appearing in eq.(\ref{IDCS}) is commonly performed with the help of FEYNCALC \cite{feyncalc1,feyncalc2,feyncalc3}. The results we obtained are included in the appendix.
\section{NUMERICAL RESULTS AND DISCUSSION}\label{NUMERICAL RESULTS}
In this part, we perform a numerical study that allows us to understand the variations of the DCS of the scattering process $ e^{-}+\nu_{\mu}\longrightarrow e^{-}+\nu_{\mu} $ either in the absence or in the presence of an external laser field of circular polarization. To evaluate the numerical values of the DCS, we have fixed the kinetic energy of the incident muon neutrino at $E_{\nu}^{\text{kin}}=0.5\times10^{-3}~\text{GeV}$ for low energy scan while $E_{\nu}^{\text{kin}}=10~\text{GeV}$ for high energy scan. In addition, the energy of the incident electron varies from $E_{e}^{\text{kin}}=10^{-3}~\text{GeV}$ towards high energy values. For the geometry of the collision, we have chosen the wave vector $\vec{k}$ to be oriented along the $z$-axis, and the incident muon neutrino $k_{i}$ moves in the inverse of $z$-axis direction. In addition, the polarization vectors $\eta_{1}$ and $\eta_{2}$ are chosen to be perpendicular to $\vec{k}$ such that $\eta_{1}$ being in the $x$-axis direction while $\eta_{2}$ is along the $y$-axis. 
Besides, both incoming and outgoing electrons are chosen in a general geometry with spherical coordinates $\theta_{i}$, $\phi_{i}$, $\theta_{f}$ and $\phi_{f}$ (figure \ref{fig1}). 
The outgoing muon neutrino momentum can be determined  by using the four-momentum conservation law $q_{f}+k_{f}-q_{i}-k_{i}-nk=0$. In this work, we consider two different geometries: The first one is chosen such that $\theta_{i} =1^{\circ}$, $\phi_{i} = \phi_{f} = 0^{\circ}$ and  $-180^{\circ}\leq\theta_f\leq 180^{\circ}$ in order to compare our results with those obtained in \cite{ElAsri}. The second choice of the geometry parameters is as follows: $\theta_{i} =\phi_{i}$, $\phi_{f} = 90+\phi_{i}$ and $-180^{\circ}\leq\theta_f\leq 180^{\circ}$. This second choice is based on the fact that it leads to coherent and good results. We note that both geometries give the value of the maximum DCS (peak) around $\theta_{f}=0^{\circ}$.
\subsection{In the absence of the laser field }
\begin{figure}[hbtp]
 \centering
\includegraphics[scale=0.44]{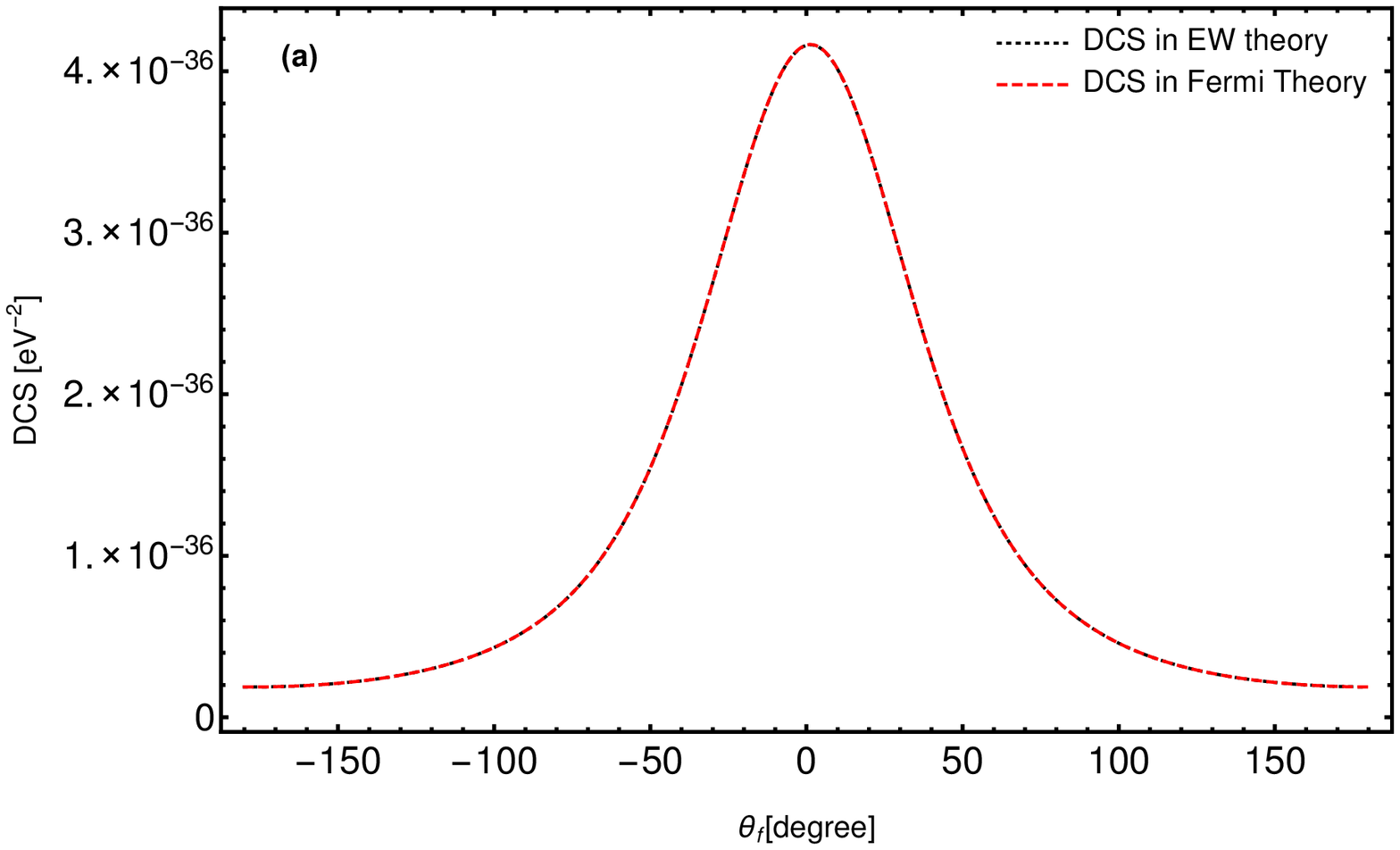}\hspace*{0.11cm}
\includegraphics[scale=0.44]{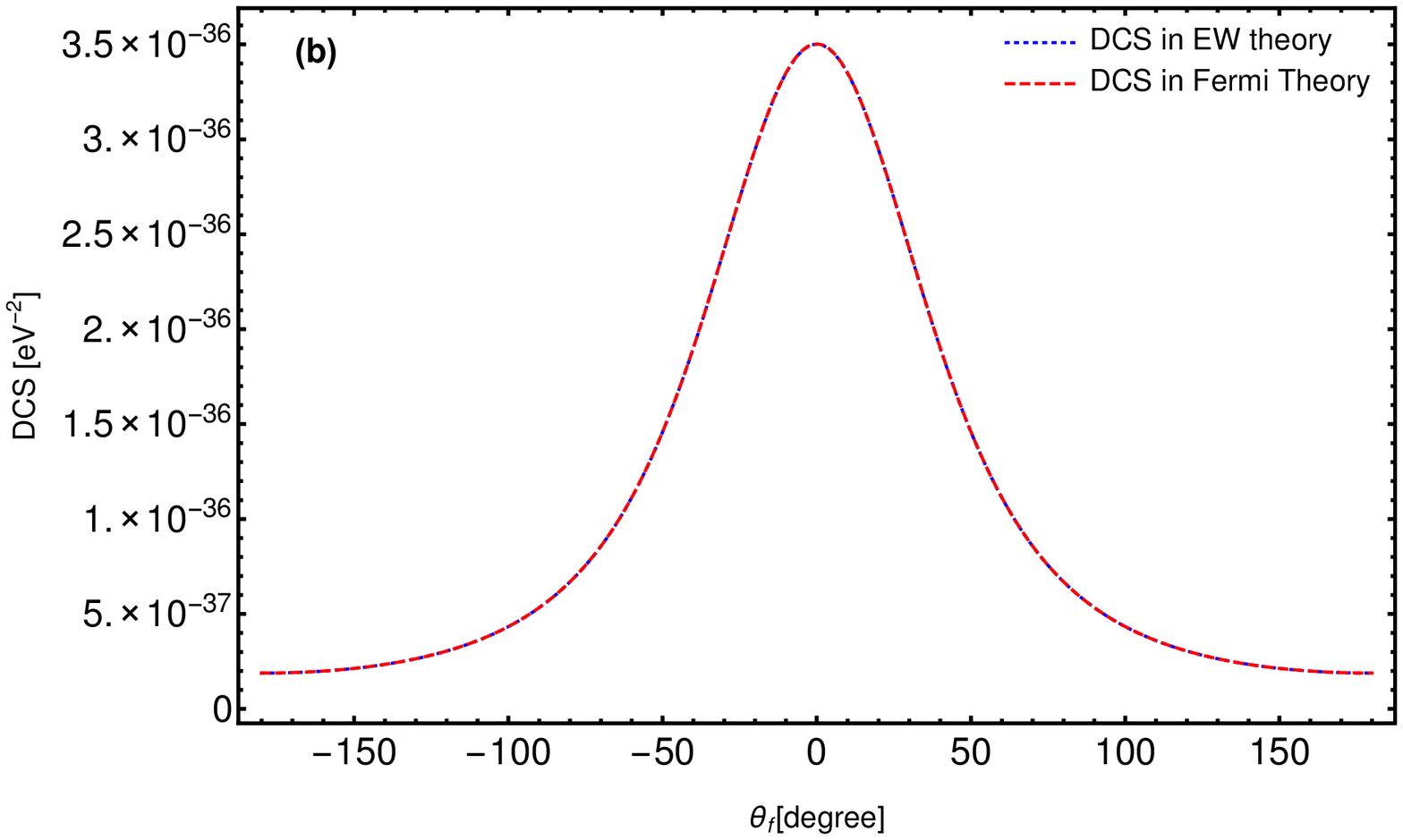}
\caption{Comparison between the variation of the DCS in Fermi theory \cite{ElAsri} and that obtained in electroweak theory (equation (\ref{freedcs})) as a function of the scattering angle $\theta_{f}$ for different geometries. The kinetic energy of the incoming muon neutrino and incident electron are $E_{\nu}^{\text{kin}}=0.5\times10^{-3}~\text{GeV}$ and  $E_{e}^{\text{kin}}=10^{-3}~\text{GeV}$, respectively. The geometry parameters are chosen as: (a) $\theta_{i}=1^{\circ},~\phi_{i}=\phi_{f}=0 $; (b) $\theta_{i}=\phi_{i}=15^{\circ}$, $\phi_{f}=105^{\circ}$.}\label{fig3}
\end{figure}

\begin{figure}[hbtp]
 \centering
\includegraphics[scale=0.35]{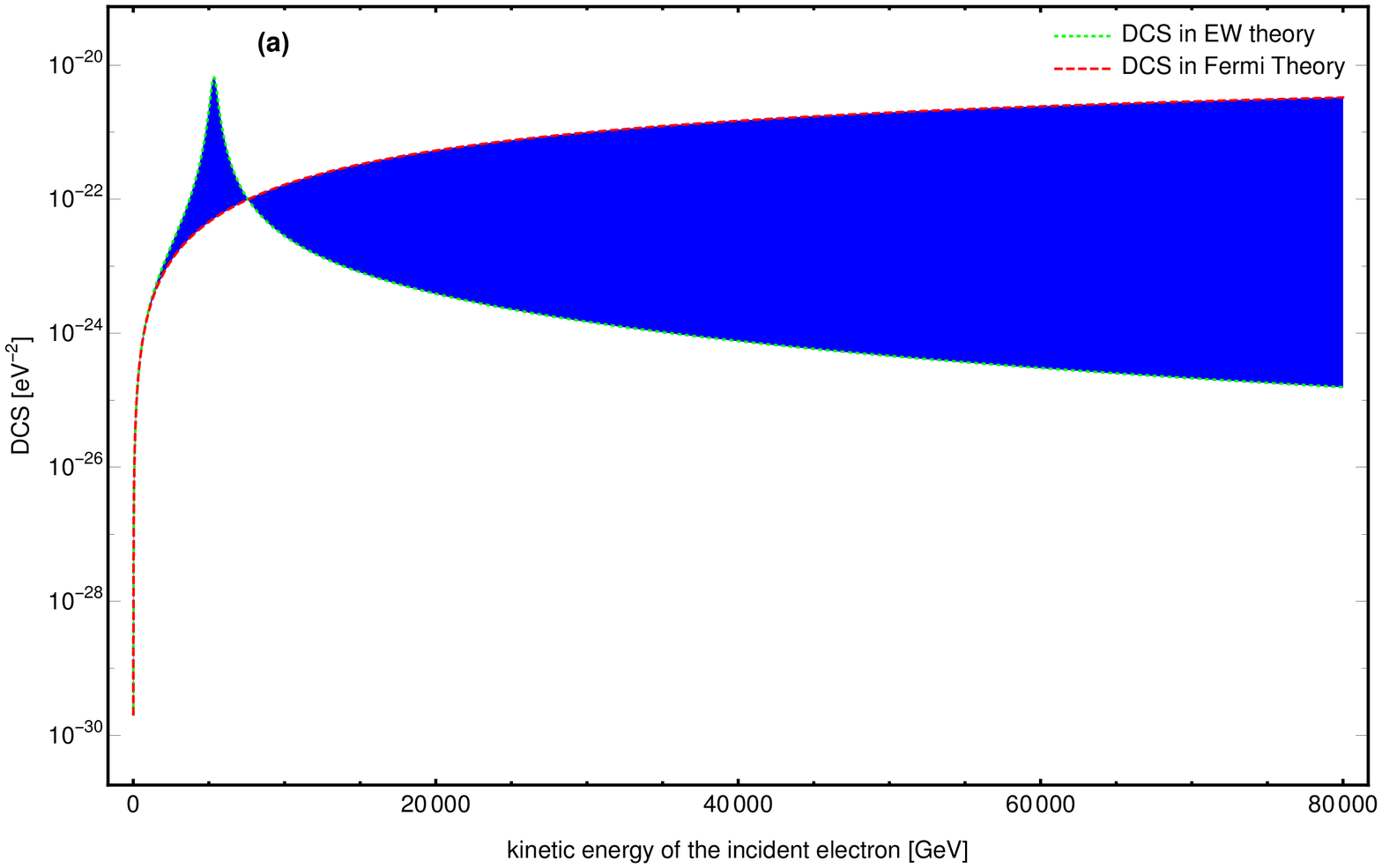}\hspace*{0.11cm}
\includegraphics[scale=0.34]{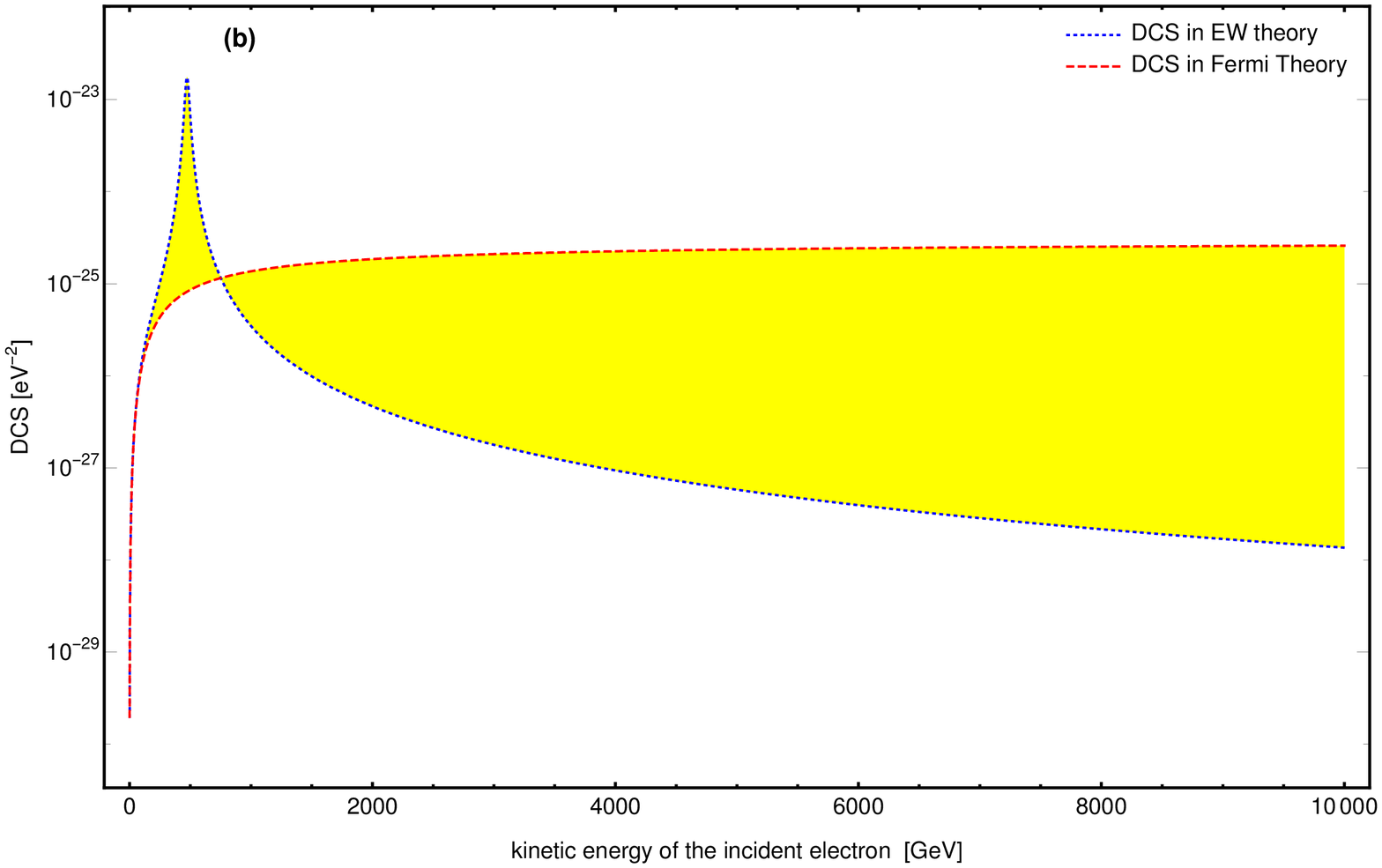}
\caption{ The laser-free DCS of the processes $e^{-} \nu_{\mu} \rightarrow e^{-} \nu_{\mu}$ in terms of the kinetic energy of the incident electron $E_{e}^{\text{kin}}$ for different geometry parameters in both electroweak theory (equation (\ref{freedcs})) and Fermi theory \cite{ElAsri}. The kinetic energy of the incident muon neutrino is taken as $ E_{\nu}^{\text{kin}}=10~GeV$. The geometry parameters are chosen as: (a) $\theta_{i}= 1^{\circ}$, $\phi_{i}=\phi_{f}=0^{\circ}$ ; (b) $\theta_{i}=\phi_{i}=15^{\circ}$, $\phi_{f}=105^{\circ}$.\label{fig4} }
\end{figure}

\begin{figure}[hbtp]
 \centering
\includegraphics[scale=0.44]{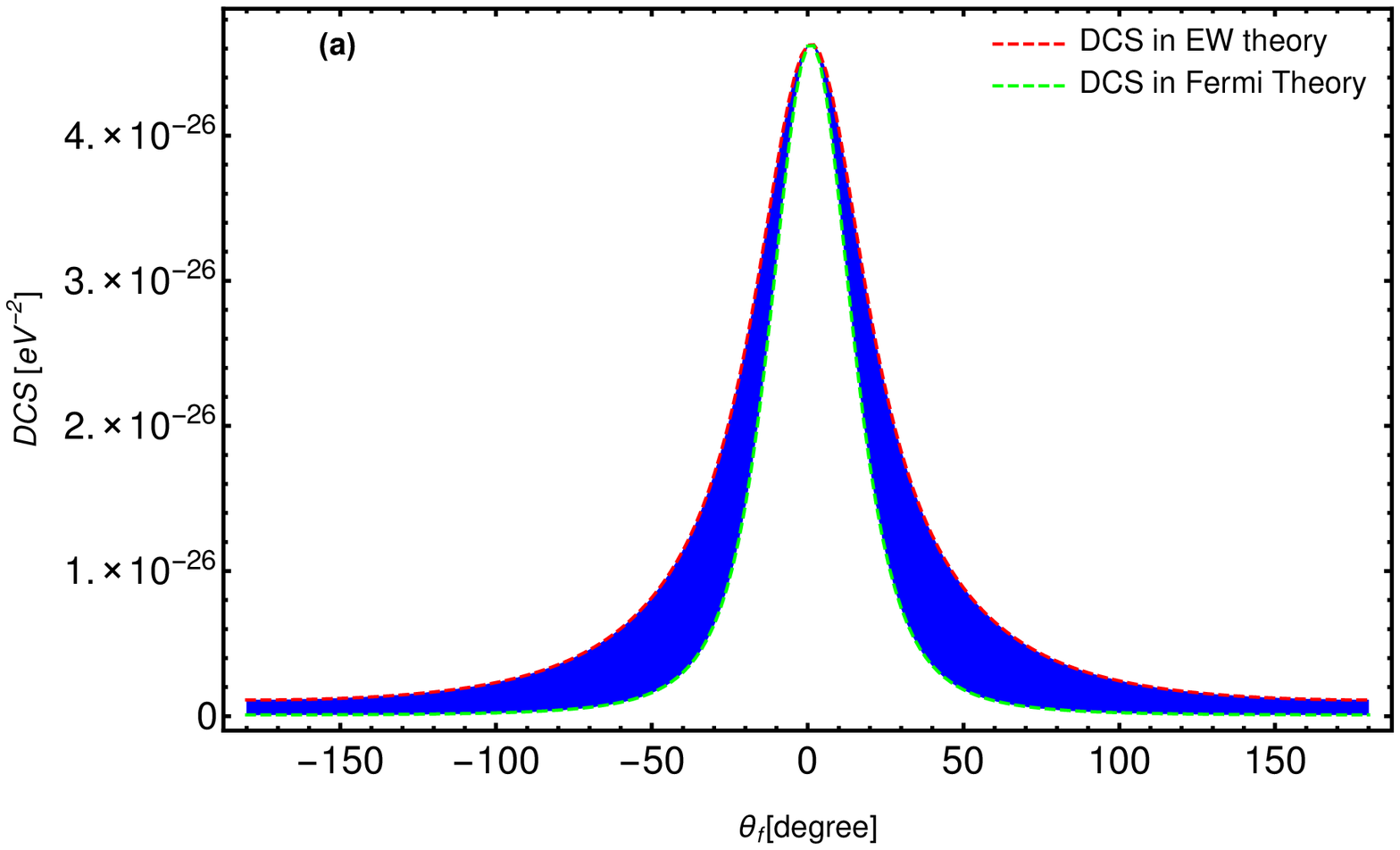}\hspace*{0.11cm}
\includegraphics[scale=0.44]{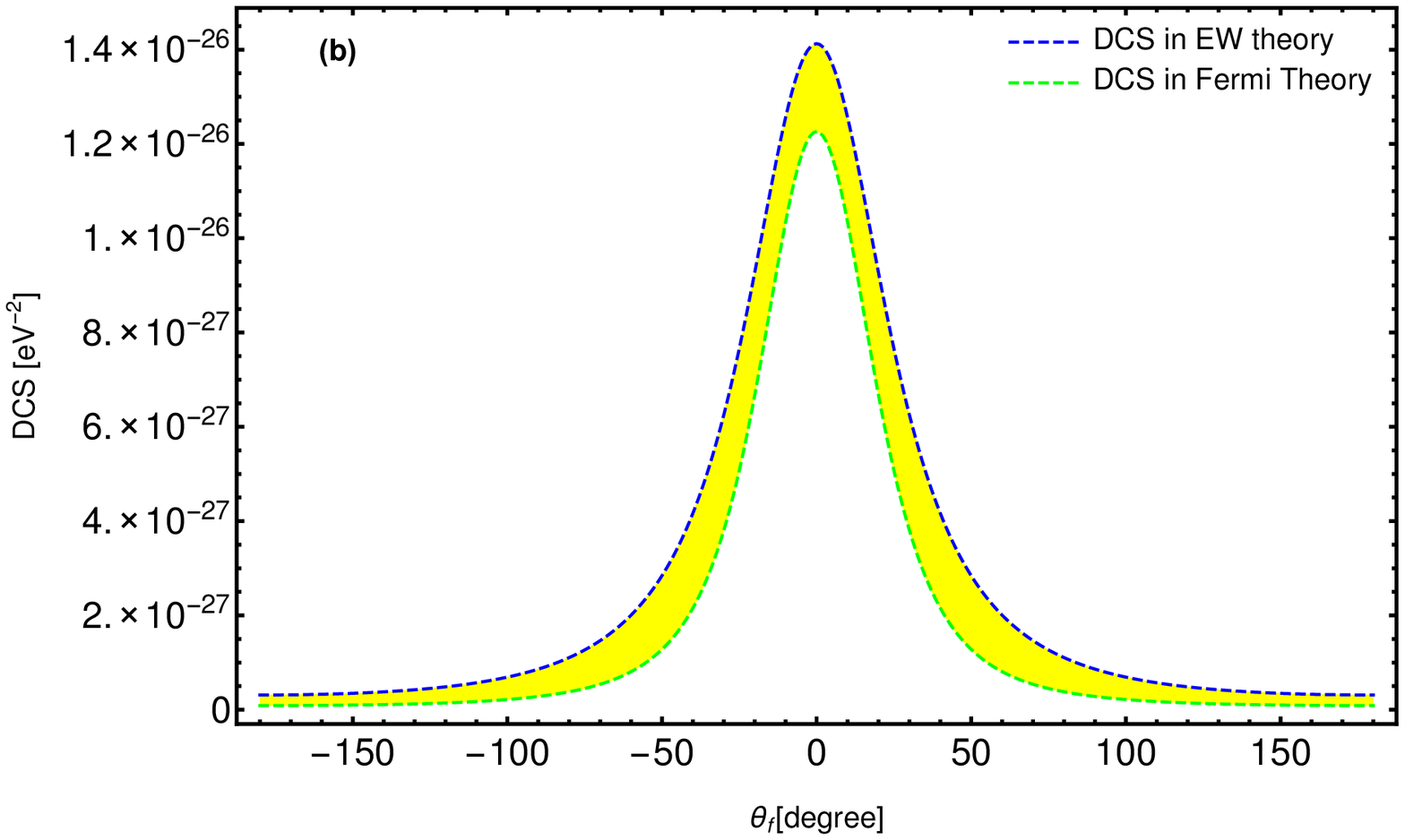}
\caption{The laser-free DCS of the elastic scattering of a muon neutrino by an electron as a function of the scattering angle $\theta_{f}$ in Fermi and electroweak theories. The kinetic energy of the incident electron is $ E_{e}^{\text{kin}}=100~\text{GeV}$, and the geometry parameters are chosen as: (a) $\theta_{i}= 1^{\circ}$, $\phi_{i}=\phi_{f}=0^{\circ}$ ; (b) $\theta_{i}=\phi_{i}=15^{\circ}$, $\phi_{f}=105^{\circ}$.}\label{fig5}
\end{figure}

\begin{figure}[hbtp]
\centering
\includegraphics[scale=0.5]{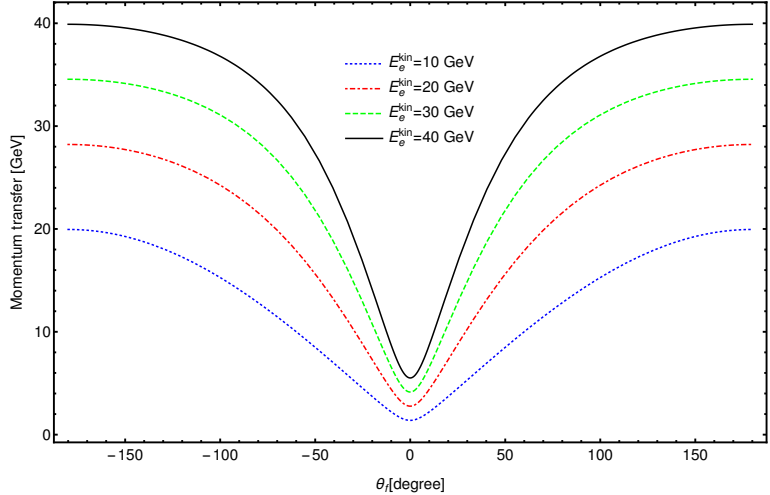}
\caption{The laser-free relativistic four-momentum transfer $q=p_{f}-p_{i}$ (eq. \ref{fourvectran}) as a function of the scattering angle $\theta_{f}$ for various kinetic energies of the incoming electron. The kinetic energy of the incident muon neutrino is $E_{\nu}^{\text{kin}}=10 ~\text{GeV}$, and the geometry parameters are taken as $\theta_{i}=\phi_{i}=8^{\circ}$ and $\phi_{f}=98^{\circ}$.}\label{fig8}
\end{figure}

In this subsection, we start our discussion by comparing our results obtained in the framework of electroweak theory with those obtained by El Asri, et al \cite{ElAsri} in the Fermi theory. 
The latter can be considered as a special case of the electroweak theory.  
In addition, it has been found that the expression of the DCS in the framework of the electroweak theory (equation (\ref{freedcs})) becomes equal to its corresponding expression in the Fermi theory whenever the relativistic laser-free momentum transfer is equal to zero. 
In this case, there is no propagator exchange between the incident and scattered particles during the collision process, i.e., the interaction is punctual. 
However, when the relativistic four-momentum transfer is non-null, the two DCSs in the framework of both Fermi and electroweak theory are identical only at low energies, and they are different if the incoming particles move with high energies.
To well understand the behavior of the DCS, it is necessary to study its variations as a function of both the final scattering angle $\theta_{f}$ and the kinetic energy of the incident electron. 
In figure \ref{fig3}, we plot the laser-free DCS of the scattering process of a muon neutrino by an electron (equation \ref{freedcs}), and we have compared it with its corresponding laser-free DCS in the Fermi theory for various geometries.
The geometry parameters are chosen in figure \ref{fig3}(a) as $ \theta_{i}=1^{\circ}$ and $\phi_{i}=\phi_{f}=0^{\circ}$  while  $ \theta_{i}=\phi_{i}=15^{\circ}$ and $ \phi_{f}=105^{\circ}$ in figure \ref{fig3}(b). 
We see that both curves take the form of a Gaussian function with a maximum value (peak) around $\theta_{f} =0^{\circ}$. 
We also notice from these figures that the DCS that corresponds to the Fermi theory coincides with that of the Electroweak theory. This result implies that both theories are valid to study low energy scattering processes, and this is in agreement with literature results \cite{greiner}. 
Let's move now to study the effect of the kinetic energy of the incoming particles on the laser-free DCS in the framework of the two theories.
In figure \ref{fig4}, we consider the dependence of the laser-free DCS on the kinetic energy of the incident electron $E_{e}^{\text{kin}}$ for different geometry parameters. 
According to this figure, we remark that the DCS in the Fermi theory increases rapidly in a very small range of low energies, and it keeps increasing slowly above this energy range.
This result is in accordance with the behavior of the unitarity violation of the $S_{fi}$ matrix. 
For this reason, the electroweak theory comes to solve this problem of unitarity violation of the DCS as it is illustrated in figures \ref{fig4}(a) and \ref{fig4}(b). 
We also remark that, the DCS in the framework of the electroweak theory increases with the increase of the kinetic energy of the incident electron until it reaches $E_{e}^{\text{kin}}=5244.19 \text{GeV}$ and $E_{e}^{\text{kin}}=456.578  \text{GeV}$ in figures \ref{fig4}(a) and \ref{fig4} (b), respectively. Then, it decreases abruptly. 
In this critical point where the DCS is maximum, we get the resonance which occurs at $q=M_{\text{Z}}$ (production of $Z$-boson) for both geometries.  
This is due to the fact that the relativistic four-momentum transfer ($q$) is a function depending on both geometry parameters and the kinetic energy of the incoming electron. Indeed, if the geometry parameters changes, the kinetic energy of the incident electron also changes. However, the resonance peak always occurs at $q=M_{\text{Z}}$. 
In summary, we have found that the two DCSs in both theories are coincident at low energies, but they are different at high energies.
To obtain more information, we have plotted in figure \ref{fig5} the two DCSs as a function of the scattering angle by using the same geometry parameters as in the previous figures. 
In this case, the kinetic energy of the incoming electron and muon neutrino are taken as $E_{e}^{\text{kin}}=100~\text{GeV}$ and $E_{\nu}^{\text{kin}}=10~\text{GeV}$, respectively. 
According to figure \ref{fig5}, we notice that, at high energies, the DCS in the framework of the electroweak theory is the most dominant compared to that of the Fermi theory, and this implies that the electroweak theory is the most valid model to study scattering processes in high energy physics. 
The small correction that appears in figure \ref{fig5} (a) around $\theta_{f}=0^{\circ}$ can be explained by the effect of geometry on the DCS. Indeed, the geometry has an important influence on the variations of the DCS of scattering processes.
We will close our discussion in this subsection by displaying, in figure \ref{fig8}, the dependence of the relativistic four-vector momentum transfer outside the laser field on the scattering angle for various kinetic energies of the incident electron. The squared relativistic four-vector momentum transfer in the absence of the laser field can be written as follows:
\begin{equation}\label{fourvectran}
q^{2}=(p_{f}-p_{i})^{2}=~2m^{2}-2~~p_{i}^{0}~p_{f}^{0}+2~|\textbf{p}_{i}|~|\textbf{p}_{f}|~F(\theta_{i},\theta_{f},\phi_{i},\phi_{f}).
\end{equation}
From this equation, we can deduce that the relativistic four-vector momentum transfer depends on both incoming particles energies and geometry parameters. Due to this dependence, the four-vector momentum transfer is considered as a fundamental physical quantity that characterizes all scattering theories. 
In addition, we remark that it is symmetrical with respect to the axis $\theta_{f}=0^{\circ}$, and it takes its minimum value around $\theta_{f}=0^{\circ}$. We remark also that it rises by increasing either or both the kinetic energy of the incoming electron and the final scattering angle. 
This behavior was clearly mirrored in the DCS obtained in the framework of the electroweak theory for that its expression (equation (\ref{freedcs})) is conversely proportional to $\big(q^{2}-M_{Z}^{2}\big)^{2}$.
As a consequence, the DCS has its maximum value around $\theta_{f} = 0^{\circ}$, and it rises by increasing the kinetic energy of the incoming electron.
\subsection{In the presence of the laser field}
\begin{figure}[hbtp]
 \centering
\includegraphics[scale=0.42]{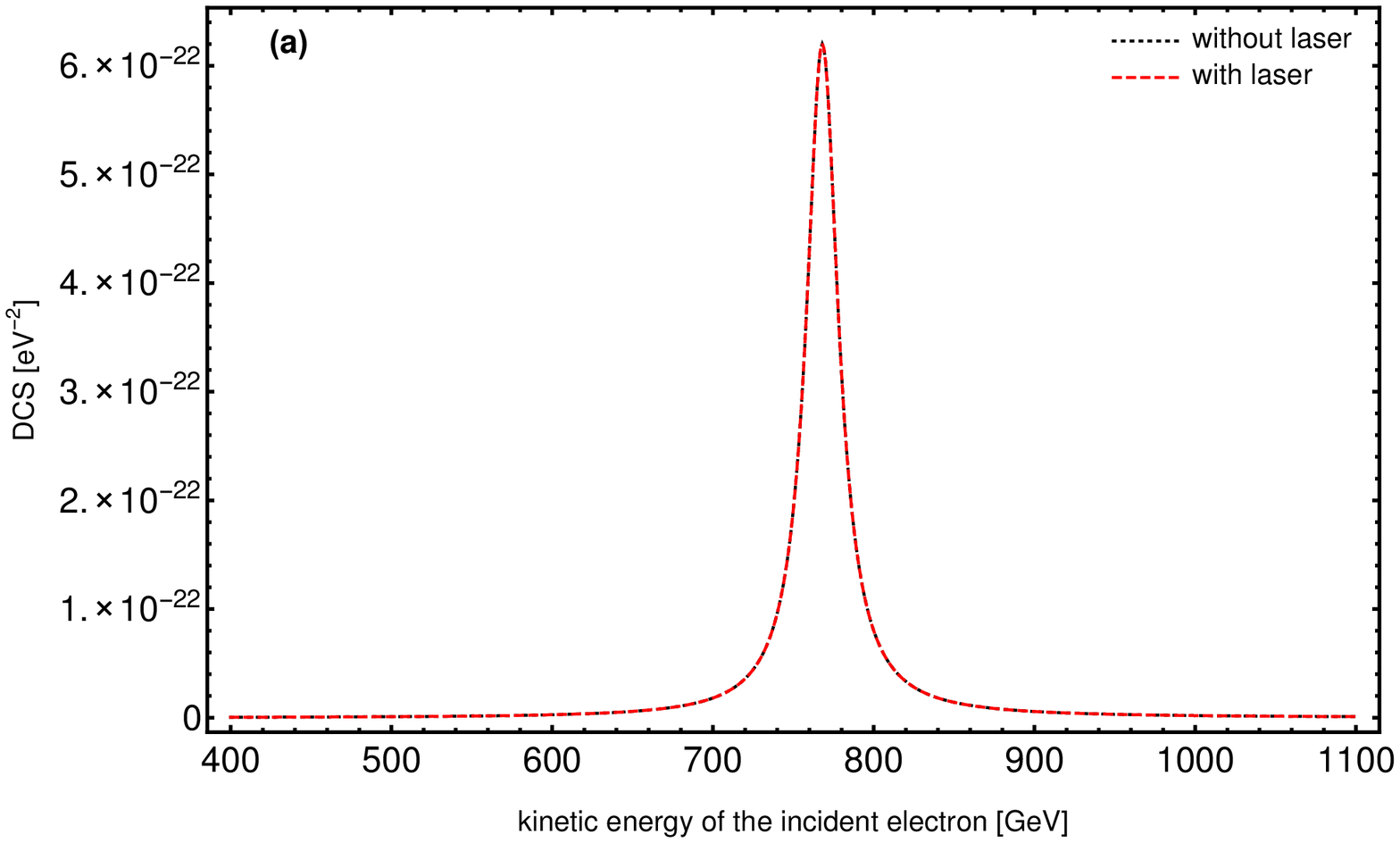}\hspace*{0.11cm}
\includegraphics[scale=0.42]{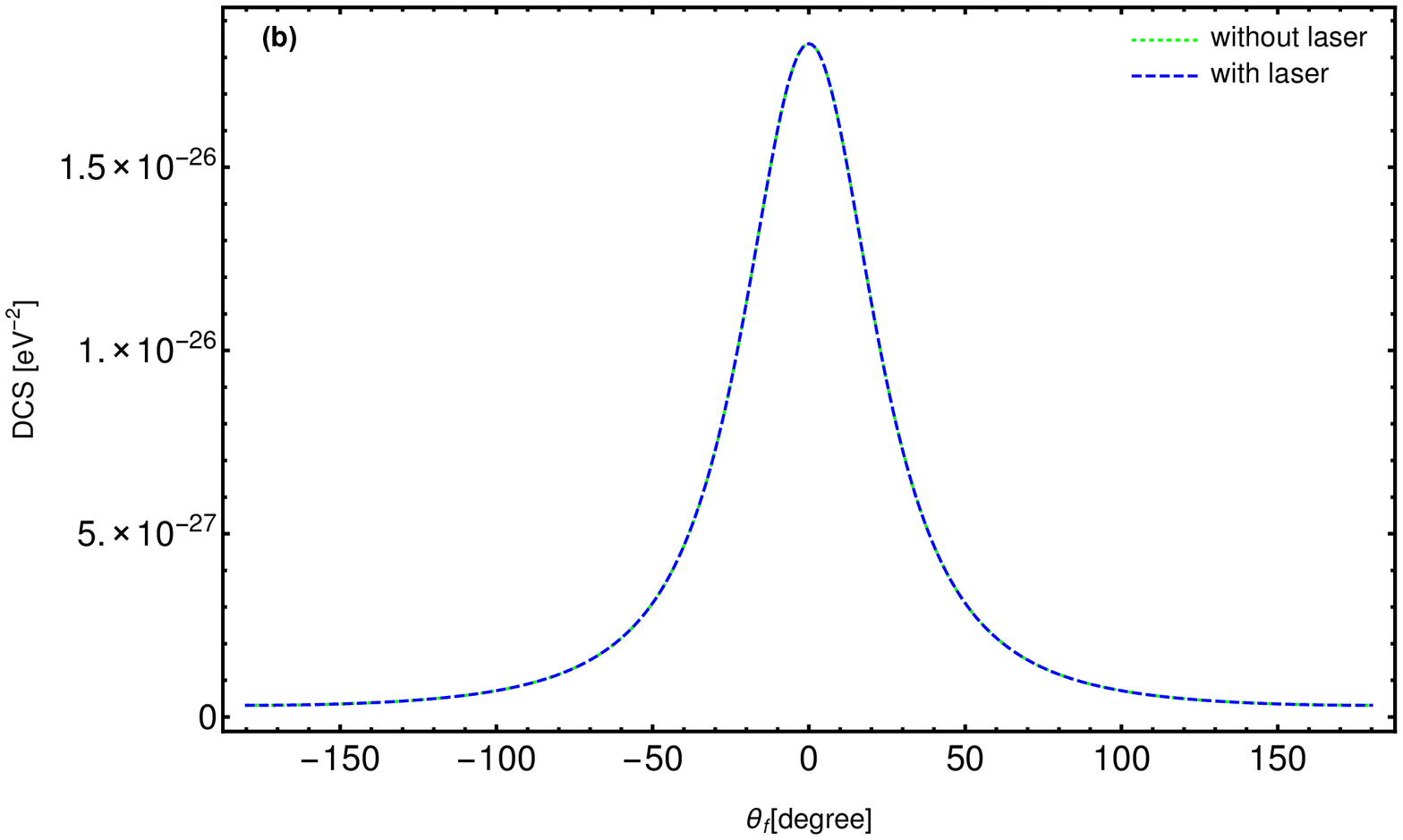}
\caption{Comparison between the DCS without laser and the DCS in the presence of a laser field by taking the electric field strength as $\xi_{0} = 0~\text{V/cm}$ and without any exchange of photons ($n = 0$). The kinetic energy of the incident muon neutrino is $E_{\nu}^{\text{kin}}=10~\text{GeV}$, and the other parameters are chosen as: (a) $\theta_{i}=\phi_{i}=8^{\circ}$, $\theta_{f}=0^{\circ}$ and $\phi_{f}=98^{\circ}$; (b) $E_{e}^{\text{kin}}=100~\text{GeV}$,  $\theta_{i}=\phi_{i}=8^{\circ}$ and $\phi_{f}=98^{\circ}$.}\label{figlaser2}
\end{figure}

\begin{figure}[h]
 \centering
\includegraphics[scale=0.4]{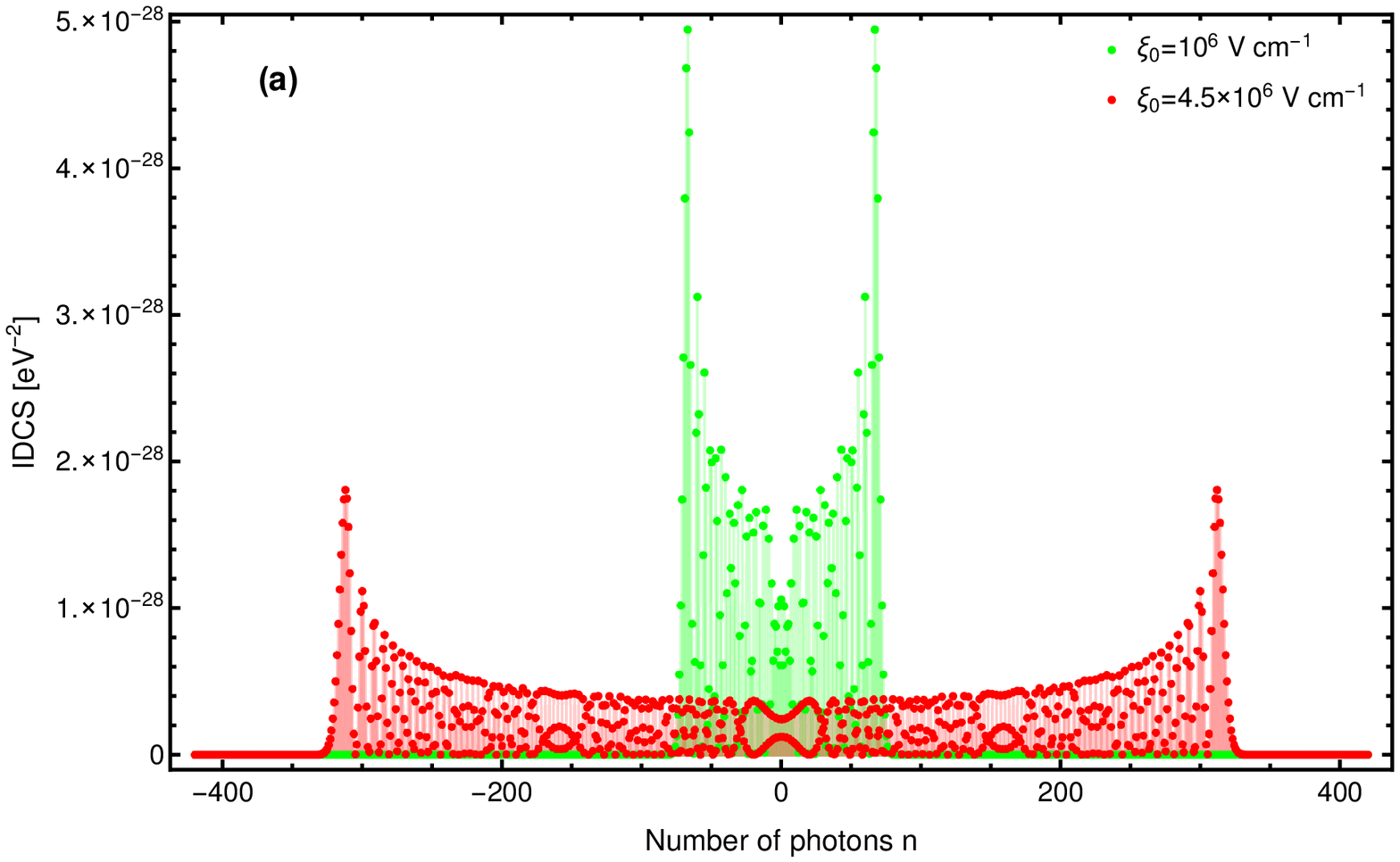}\hspace*{0.11cm}
\includegraphics[scale=0.34]{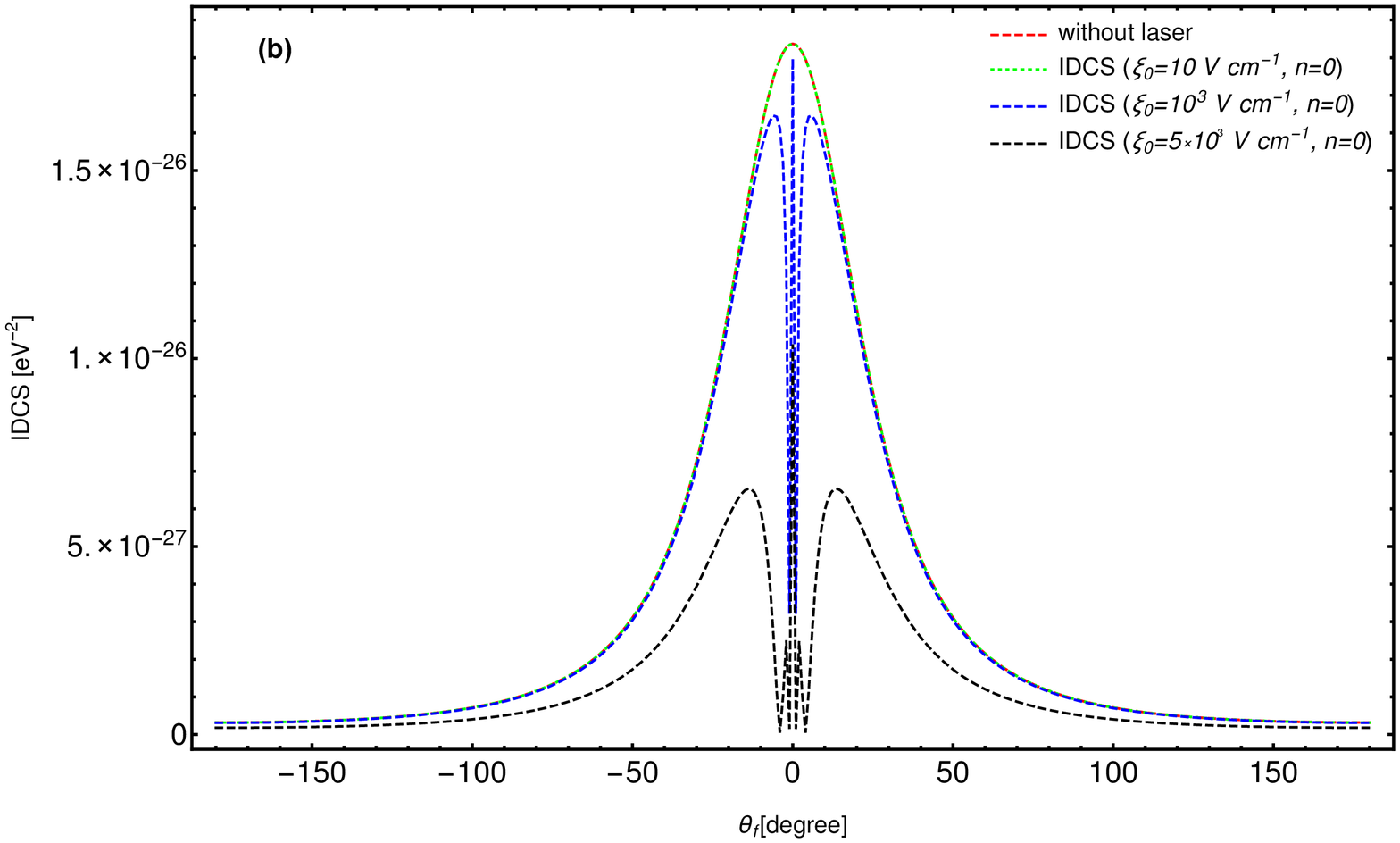}
\caption{Laser-assisted individual DCS in terms of the number of exchanged photons (a) or scattering angle $\theta_{f}$ (b) for various laser field strengths and by taking $n = 0$ in (b). The other parameters are taken as follows: $E_{\nu}^{\text{kin}}=10~\text{GeV}$, $E_{e}^{\text{kin}}=100~\text{GeV}$, $\theta_{i}=\phi_{i}=8^{\circ}$, $\theta_{f}=0^{\circ}$ and  $\phi_{f}=98^{\circ}$.}\label{figlaser4}
\end{figure}

In this part, we will present and discuss the numerical results concerning the laser-assisted DCS of the scattering process ($e^{-}+\nu_{\mu}\longrightarrow e^{-}+\nu_{\mu}$) in the framework of the electroweak theory. 
The parameters that appear when we immerse charged particles in an external electromagnetic field are the strength $(\xi_{0})$, the frequency and the number of photons exchanged between the laser field and the colliding system. 
The strength and frequency are the parameters that characterize the laser field, whereas, the number of exchanged photons appears thanks to the introduction of ordinary Bessel functions.
In addition, particles inside a laser field may change their properties, such as their effective mass and their four-momentum, and they acquire new ones. 
Thus, by including the scattering process (\ref{process}) in an external laser field, its DCS becomes dependent on both the laser-free parameters and other parameters that appear while the laser field is present. 
As a consequence, the DCS changes significantly in the presence of the laser field. 
Now, we will see what happened in the case where no photon is exchanged between the laser field and the scattering process (n=0), and the laser field strength is chosen as $0\, V.cm^{-1}$.
In this case, the argument of the ordinary Bessel functions in equation (\ref{argument}) is equal to zero ($z=0$). 
So, all the coefficients that contribute to the spinorial part of the laser-assisted DCS (equation \ref{appendix2}) will be zero except the coefficient $\Delta_{1}$ multiplied by the Bessel function $b_{n}(z)$. 
In this limit, we have $b_{n}(z)=1$ and $|\textbf{a}|=0$. Therefore, we obtain the expression of the spinorial part given by the equation (\ref{appendix1}) in the absence of the laser field. 
Concerning the other quantities that constitute the DSC in equation (\ref{dcswithlaser}), we obtain their expressions in the absence of the laser field by taking $\xi_{0} = 0~\text{V/cm}$ and $n=0$ . 
Figure \ref{figlaser2} illustrates the comparison between the laser-assisted DCS of the scattering of a muon neutrino by an electron (equation\ref{dcswithlaser}) and its corresponding DCS in the absence of the laser field.  
This comparison allows us to verify numerically our results by taking the limit of the laser field strength $\xi_{0}=0~\text{V/cm}$ and the number of photons exchanged $n=0$.
We remark that the two graphs illustrated in the two figures \ref{figlaser2} (a) and (b) are in full agreement as they are indistinguishable at all kinetic energies of the incoming electron and for all final scattering angles. 
Hence, by taking the laser parameters as zero, the laser-assisted DCS is exactly equal to its corresponding one in the absence of the electromagnetic field. 
We move now to study the DCS in the case where the electromagnetic field parameters are not equal to zero. We present, in Table \ref{table1}, the numerical values of the individual DCS of the scattering process $e^{-}+\nu_{\mu}\longrightarrow e^{-}+\nu_{\mu}$ versus the number of photons exchanged for various laser frequencies.
We observe that, by changing the laser instrument (varying the laser frequency), the number of transferred photons changes. Consequently, the individual DCS also changes. In addition, when the number of photons exchanged rises, this individual DCS varies progressively until it falls abruptly to zero. 
This result allows us to determine the "cutoff", where the individual DCS is null. 
This cutoff can be explained through the properties of the ordinary Bessel functions, which drop abruptly when their order approaches their argument. We can also see that the individual DCS is symmetrical with respect to the  $n=0$ axis, which implies that the processes of emission ($\text{n} > 0$) and absorption ($\text{n} < 0$) of photons are exactly identical. 
We also note that the number of exchanged photons interval $[-\text{n},+\text{n}]$ becomes larger by decreasing the laser field frequency. Concerning the order of magnitude of the DCS, we remark that it increases by increasing the laser field frequency.
\begin{table}[h]
\centering
\caption{Numerical values of  $d\overline{\sigma}^{n}/d\Omega$ (\ref{IDCS})~in terms of the number of photons exchanged $n$ for different frequencies of the laser field. The kinetic energies of the incident muon neutrino and the electron are successively $E_{\nu}^{\text{kin}}= 10~\text{GeV}$ and  $E_{e}^{\text{kin}}=100~\text{GeV}$, and the laser field strength is taken as $\xi_{0}=10^{6}~\text{V/cm}$. The geometric parameters are chosen as follows:  $\theta_{i}=\phi_{i}=15^{\circ}$, $\theta_{f}=0^{\circ}$ and $\phi_{f}=105^{\circ}$.}\label{table1}
\begin{tabular}{|p{1cm}|p{3.4cm}|p{1cm}|p{3.4cm}|p{1.5cm}|p{3.4cm}|}
 \hline
~~$n$ &~~ \shortstack{ $d\overline{\sigma}^{n}/d\Omega$ [eV$^{-2}$]\\He-Ne laser \\
($\hbar\omega=2~\text{eV}$)}  & ~~$n$ &~~ \shortstack{$d\overline{\sigma}^{n}/d\Omega$ [eV$^{-2}$]\\Nd:YAG laser\\
 ($\hbar\omega=1.17~\text{eV}$)} & ~~$n$ & ~~ \shortstack{ $d\overline{\sigma}^{n}/d\Omega$ [eV$^{-2}$]\\$\text{CO}_{2}$ laser \\
  ($\hbar\omega=0.117~\text{eV}$)} \\
 \hline
$\pm$~120   & 0    &$\pm$~400&  0 & $\pm$~ 12000& 0\\
$\pm$~100   & 0    &$\pm$~300& 0 & $\pm$~11500& 0\\
$\pm$~80   & $1.570\times10^{-64}$    &$\pm$~200&  $1.637 \times10^{-94}$ & $\pm$~8000& $8.510\times10^{-32}$\\
$\pm$~60   & $5.826\times10^{-43}$    &$\pm$~100&  $1.380 \times10^{-29}$  & $\pm$~4000& $3.507\times10^{-31}$\\
$\pm$~50   & $9.658\times10^{-35}$    &$\pm$~80&  $1.001 \times10^{-29}$  & $\mp$~1000& $3.042\times10^{-32}$\\
$\pm$~40   & $4.046\times10^{-29}$    &$\pm$~60&  $9.778\times10^{-29}$  & $\mp$~500& $1.700\times10^{-31}$\\
$\pm$~30   & $2.050\times10^{-28}$     &$\pm$~40&  $ 8.648\times10^{-29}$  & $\pm$~100&$6.284\times10^{-31}$\\
$\pm$~20   & $1.814\times10^{-29}$    &$\pm$~20&  $5.828 \times10^{-29}$  & $\mp$~60& $7.845\times10^{-31}$\\
$\pm$~10   & $2.236\times10^{-28}$     &$\pm$~10& $ 4.066\times10^{-29}$  & $\mp$~10& $8.190\times10^{-31}$\\
$\pm$~6   & $1.798\times10^{-28}$     &$\pm$~6&  $1.793 \times10^{-29}$  & $\pm$~6& $8.193\times10^{-31}$\\
~0   & $6.646\times10^{-29}$     &~0&  $8.195 \times10^{-30}$  & 0& $8.194\times10^{-31}$\\
 \hline
\end{tabular}
\end{table}
To more accurately visualize these results, we have inserted the figure \ref{figlaser4} that shows the variations of the individual DCS as a function of the number of photons exchanged (fig.\ref{figlaser4}(a)) and the scattering angle $\theta_{f}$ (fig.\ref{figlaser4}(b)) for different laser field strengths. 
From figure \ref{figlaser4}(a), we can observe that the electron exchanges many photons with the laser field at high strengths ($\xi_{0}=4.5 \times10^{6}~\text{V/cm}$), and the cutoff number is about ( $n = \pm~334$). However, at low laser field strengths ($\xi_{0}=10^{6}~\text{V/cm}$), the cutoff number is approximately equal to $n = \pm~100$. 
This implies that the influence of the laser field on the scattering process is more important at high field strengths, i.e., the electron heavily interacts with the laser field. According to figure \ref{figlaser4}(b), we can see that even if there is not any exchange of photons between the laser field and the colliding system, the laser field affects the DCS. 
This effect diminishes by decreasing the strength of the laser field until it vanishes. The peaks of the individual DCS as well as its oscillations and abrupt falling on the sides of figure \ref{figlaser4}(a) are affected by the well-known properties and behavior of Bessel functions.
\begin{figure}[h]
 \centering
\includegraphics[scale=0.40]{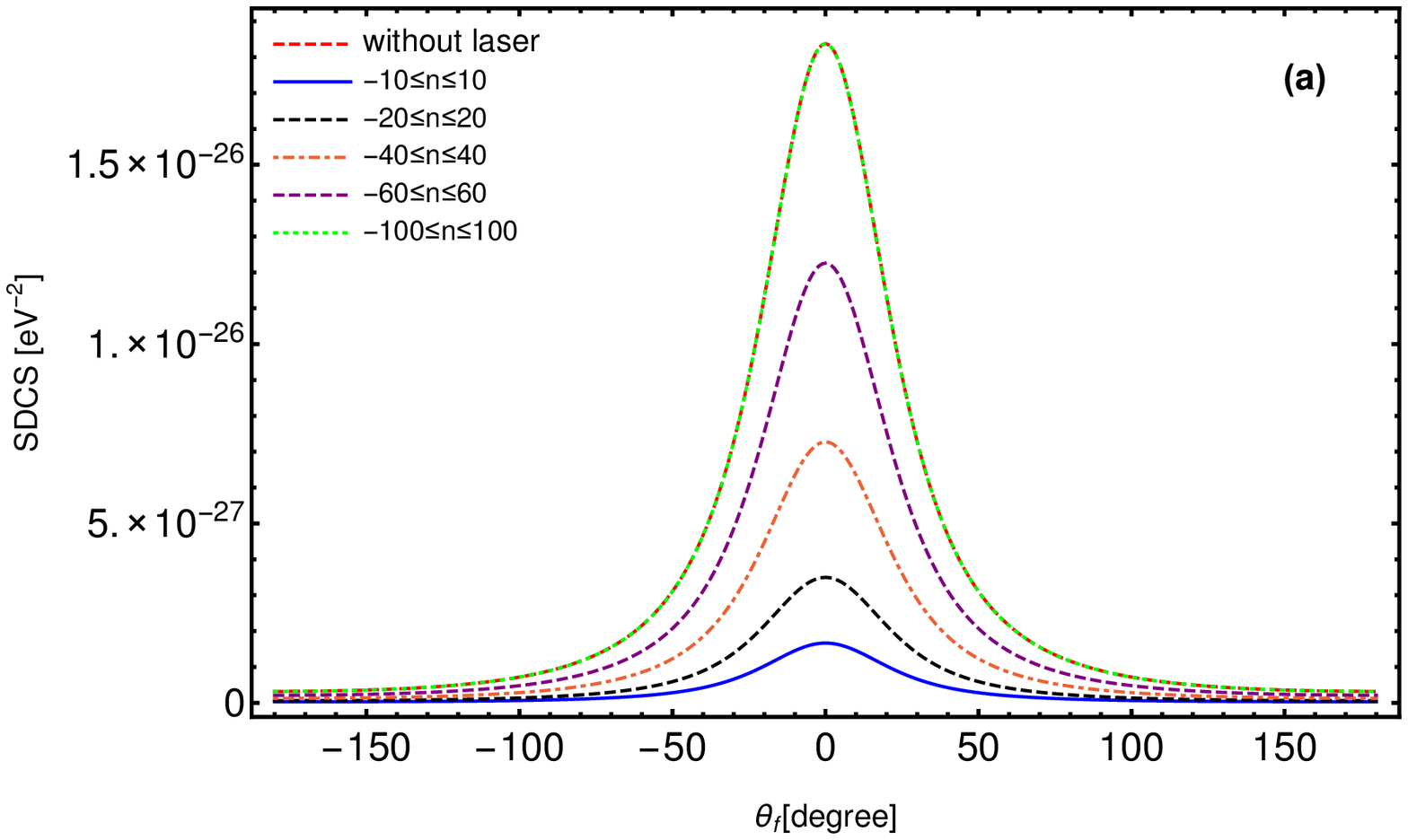}\hspace*{0.11cm}
\includegraphics[scale=0.46]{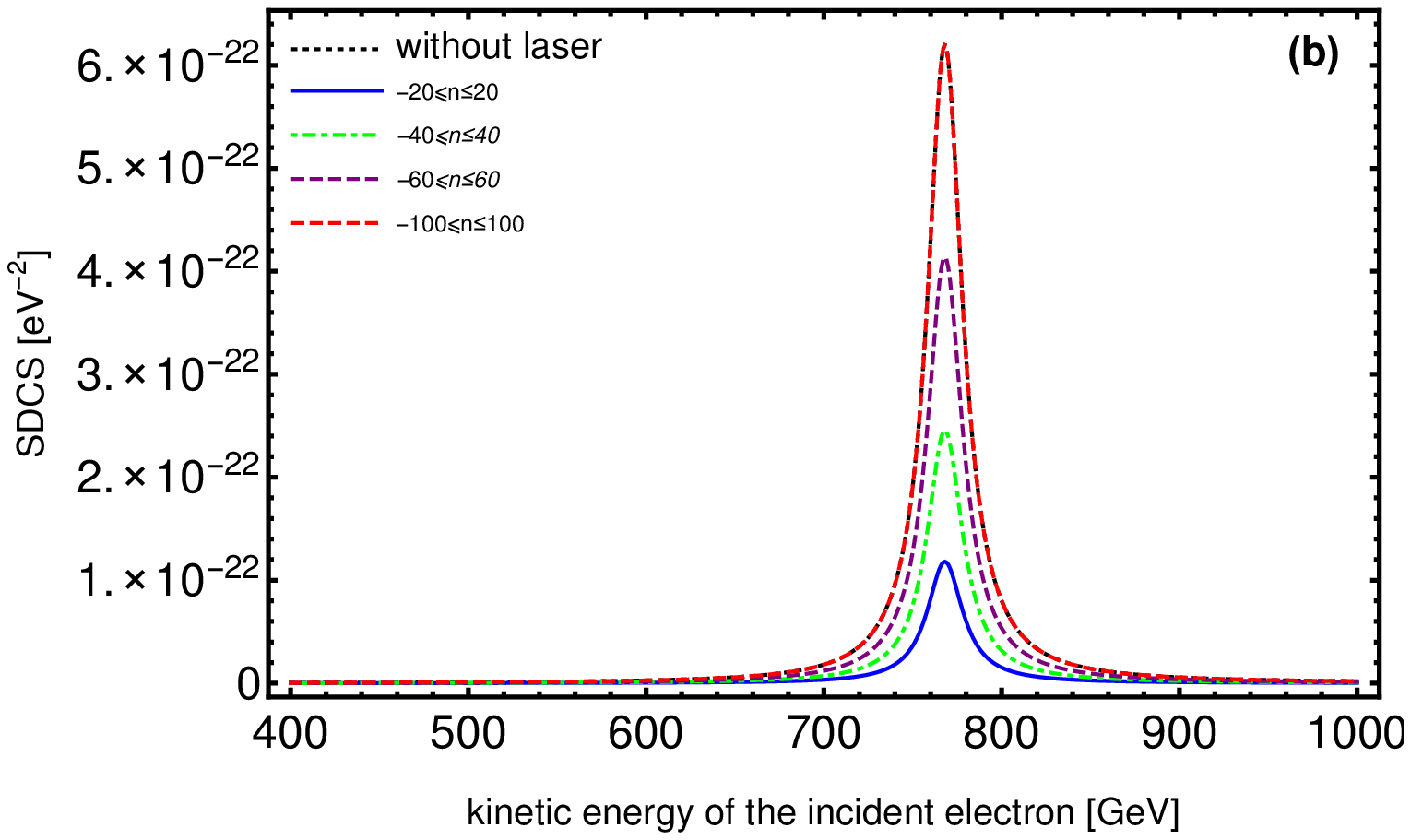}
\caption{Variations of the summed DCS of the processes $e^{-} \nu_{\mu} \rightarrow e^{-} \nu_{\mu}$ in terms of the scattering angle $\theta_{f}$ and the incident electron energy  in the presence of the laser field for different numbers of exchanged photons. The laser field parameters are chosen as $\omega=2~\text{eV}$ and $\xi_{0}=10^{6}~\text{V/cm}$. The other parameters are taken as: (a) $E_{\nu}^{\text{kin}}=10~\text{GeV}$, $E_{e}^{\text{kin}}=100~\text{GeV}$, $\theta_{i}=\phi_{i}=8^{\circ}$ and $\phi_{f}=98^{\circ}$ ; (b) $E_{\nu}^{\text{kin}}=10~\text{GeV}$, $\theta_{i}=\phi_{i}=8^{\circ}$, $\theta_{f}=0^{\circ}$  and $\phi_{f}=98^{\circ}$.   }\label{figlaser6}
\end{figure}
After this thorough discussion about the effect of the electromagnetic field-related parameters on the individual DCS, let's move on to discuss the effect of the electromagnetic fields on the summed DCS as it is an important physical quantity.
We show, in figure \ref{figlaser6}, the variation of the summed DCS of the scattering process $e^{-}+\nu_{\mu}\longrightarrow e^{-}+\nu_{\mu}$ as a function of the scattering angle $\theta_{f}$ and the kinetic energy of the incoming electron for various summations over the number of photons exchanged.
The numbers of exchanged photons that we have chosen in both figures \ref{figlaser6}(a) and \ref{figlaser6}(b) are $-N\leq n\leq+N$, with $N=10, 20, 40, 60, 100$. 
According to figure \ref{figlaser6}, we notice that the graph of the DCS in the absence of the external field is always the highest compared to that of the laser-assisted DCS. 
We also notice that when we sum over the cutoffs ($n=\pm 100$), we find that the laser-assisted summed DCS is perfectly consistent with the DCS without laser.
So, it is impossible to distinguish them from each other in all kinetic energies of the incoming electron and for all  final scattering angles $\theta_{f}$. 
This implies that the influence of the laser field on the incoming and outgoing electrons is completely canceled. In this case, the equivalence of the two DCSs with and without laser is called the sum rule and was first proposed by Kroll and Watson \cite{Watson}, and it is illustrated mathematically as follows: 
\begin{equation}
\sum_{n=-\text{cutoff}}^{+\text{cutoff}}\dfrac{d\overline{\sigma}^{n}}{d\Omega}=\bigg(\dfrac{d\overline{\sigma}}{d\Omega}\bigg)^{\text{non-laser} }.
\end{equation}
For more details and explanations about the sum-rule, see \cite{ElAsri}. 
\begin{figure}[hbtp]
\centering
\includegraphics[scale=0.52]{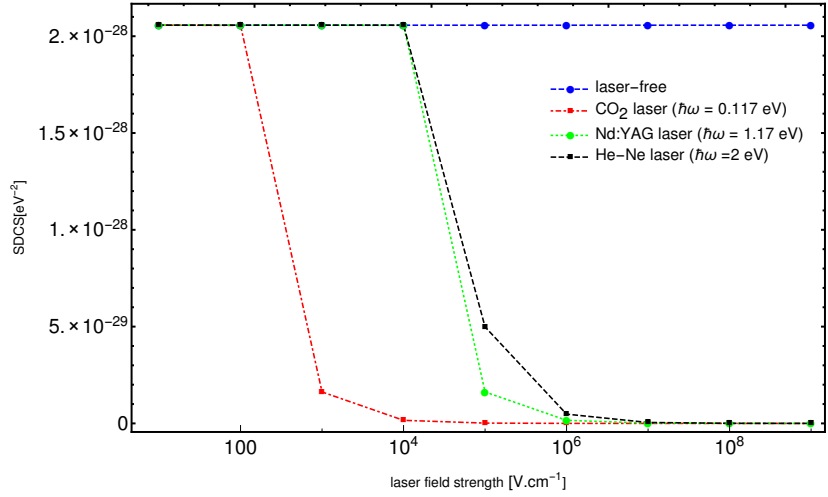}
\caption{Variation of the summed DCS (equation \ref{dcswithlaser}) in terms of the laser field strength at various frequencies. the number of photons exchanged is taken as $-20 \leq n \leq 20$. The kinetic energy of the incident muon neutrino and that of the electron are $E_{\nu}^{\text{kin}}= E_{e}^{\text{kin}}=10~\text{GeV}$. The geometry parameters are $\theta_{i}=1^{\circ}$ and $\phi_{i}=\phi_{f}=\theta_{f}=0^{\circ}$.}\label{figlaser7} 
\end{figure}
To investigate the effect of the laser parameters on the summed DCS of the scattering process (\ref{process}), we have plotted in Figure \ref{figlaser7} its variation as a function of the laser field strength for various frequencies which are $\hbar\omega=0.117~\text{eV}$ ($\text{CO}_{2}$ laser), $\hbar\omega=1.17~\text{eV}$ (Nd:YAG laser) and $\hbar\omega=2~\text{eV}$ (He-Ne laser). It can be seen, from this figure, that as the frequency is raised, the effect of the laser field on the summed DCS progressively decreases. 
For example, the $\text{CO}_{2}$ laser ($\hbar\omega=0.117~\text{eV}$) starts its effect when the laser field strength is higher than the threshold $10^{2}~\text{V/cm}$ while the He-Ne laser ($\hbar\omega=2~\text{eV}$) begins its effect from $\xi_{0}=10^{5}~\text{V/cm}$. Thus, the effect of the laser field on the summed DCS is more stronger at either or both low frequencies and high laser field strengths $\xi_{0}$.
\section{Conclusion}\label{Conclusion}
In this paper, we have performed the analytical calculation of the DCS of the elastic scattering process $e^{-}+\nu_{\mu}\longrightarrow e^{-}+\nu_{\mu}$ according to the electroweak theory in both absence and presence of a circularly polarized electromagnetic field.
In the absence of the external electromagnetic field, we have compared our results with those obtained by El Asri et al \cite{ElAsri} in the framework of the Fermi theory, and we have found that the differential cross-sections of the two theories are in excellent agreement at low energies and different at high energies. 
In the presence of the electromagnetic field, the theoretical results show that the colliding system can exchange a large number of photons with the laser field, and we have also found that the DCS of this process is modified significantly by the laser field. 
This effect of the laser on the DCS decreases at low intensities and high frequencies. In general, we can conclude that the DCS of this process can be influenced by the electromagnetic field as long as the number of photons exchanged between the colliding system and the laser field does not reach the cutoff number. 
Once the sum is made over the cutoff number, the sum rule is verified, and in this case, the influence of the laser on the DCS will be canceled. We have explained the changes of the DCS in the presence of an electromagnetic field by the behavior and properties of the Bessel functions.
\appendix
\section{Calculation of traces}\label{appendix}
To provide the expression of the quantity $\sum_{s'_{i,f},s_{i,f}}|M_{fi}|^{2}$ in equation (\ref{freedcs}), we have used the FeynCalc-9.3.0 package. Therefore, we have:
\begin{equation}\label{appendix1}
\begin{split}
 \sum_{s'_{i,f},s_{i,f}}|M_{fi}|^{2}=&\text{Tr}\big[(\slashed{p}_{f}+m)\gamma^{\mu}(g_{V}-g_{A}\gamma^{5})(\slashed{p}_{i}+m)\gamma^{\nu}(g_{V}-g_{A}\gamma^{5})\big]\\&\times \text{Tr}\big[(\slashed{k}_{f}+m_{\nu_{\mu}})\gamma_{\mu}(1-\gamma_{5})(\slashed{k}_{i}+m_{\nu_{\mu}})\gamma_{\nu}(1-\gamma_{5})\big]\\
 =& 64 \big[g_A^2 \big((k_f.p_f)(k_i.p_i)+(k_f.p_i)(k_i.p_f)+(k_i.k_f)m^2\big)+2g_A g_V \big((k_f.p_f)(k_i.p_i)\\&-(k_f.p_i)(k_i.p_f)\big)+g_V^2 \big((k_f.p_f)(k_i.p_i)+(k_f.p_i)(k_i.p_f)-(k_i.k_f)m^2\big)\big].
\end{split}
\end{equation}
The numerical evaluation of the two traces mentioned in eq. (\ref{IDCS}) gives the following result:
\begin{equation}\label{appendix2}
\begin{split}
\sum_{s'_{i,f},s_{i,f}}  |\mathcal{M}_{fi}^n|^2 =&\Delta_{1}|b_{n}(z)|^{2}+\Delta_{2}|b_{1n}(z)|^{2}+\Delta_{3}|b_{2n}(z)|^{2}+\Delta_{4}b_{n}(z)b^{*}_{1n}(z)+\Delta_{5}b_{1n}(z)b^{*}_{n}(z)\\
&+\Delta_{6}b_{n}(z)b^{*}_{2n}(z)+\Delta_{7}b_{2n}(z)b^{*}_{n}(z)+\Delta_{8}b_{1n}(z)b^{*}_{2n}(z)+\Delta_{9}b_{2n}(z)b^{*}_{1n}(z),
\end{split}
\end{equation}
where $z$ is the argument of the Bessel function defined formerly in eq. (\ref{argument}), and its order n is interpreted as the number of photons exchanged between the colliding system and the electromagnetic field. 
The five coefficients $\Delta_{5}$, $\Delta_{6}$ and $\Delta_{7}$, $\Delta_{8}$ and $\Delta_{9}$ are too long to be presented here, and the four coefficients $\Delta_{1}$, $\Delta_{2}$ $\Delta_{3}$ and $\Delta_{4}$ are expressed explicitly in terms of various scalar products as follows:
\begin{equation}
\begin{split}
\Delta_{1}=&\frac{32}{(k.p_{f}) (k.p_{i})}\bigg[|\textbf{a}|^4 e^4(g_{A}^2 + g_{V}^2)(k.k_{f})(k.k_{i})+2 (k.p_{f})(k.p_{i})\big(2 g_{A} g_{V} (-(k_{f}.p_{i})(k_{i}.p_{f})\\
&+(k_{f}.p_{f})(k_{i}.p_{i}))+g_{V}^2 ((k_{f}.p_{i})(k_{i}.p_{f})+(k_{f}.p_{f})(k_{i}.p_{i})-(k_{i}.k_{f}) m^2) \\
&+g_{A}^2 ((k_{f}.p_{i})(k_{i}.p_{f})+(k_{f}.p_{f})(k_{i}.p_{i})+(k_{i}.k_{f}) m^2)\big)+|\textbf{a}|^2 e^2 \big(2 g_{A} g_{V}\big(-(k_{i}.p_{i})\\
&\times(k.k_{f})(k.p_{f})+(k_{f}.p_{i})(k.k_{i})(k.p_{f})+(k_{i}.p_{f})(k.k_{f})(k.p_{i}) -(k_{f}.p_{f})(k.k_{i})(k.p_{i})\big)\\
&+g_{A}^2 \big((k_{i}.p_{i})(k.k_{f})(k.p_{f})+(k_{f}.p_{i})(k.k_{i})(k.p_{f})+ (k_{i}.p_{f})(k.k_{f})(k.p_{i}) \\
&+(k_{f}.p_{f})(k.k_{i})(k.p_{i})-2 (k_{i}.k_{f})(k.p_{f})(k.p_{i})-2(k.k_{f})(k.k_{i}) m^2 - 2(k.k_{f}) (k.k_{i})\\
&\times(p_{i}.p_{f})\big)g_{V}^2 \big((k_{i}.p_{i})(k.k_{f})(k.p_{f})+ (k_{f}.p_{i})(k.k_{i})(k.p_{f})+ (k_{i}.p_{f})(k.k_{f})(k.p_{i})\\
&+ (k_{f}.p_{f})(k.k_{i})(k.p_{i})-2 (k_{i}.k_{f})(k.p_{f})(k.p_{i})+2(k.k_{f}) (k.k_{i}) m^2 - 2 (k.k_{f})(k.k_{i})(p_{i}.p_{f})\big)\big)\bigg],
\end{split}
\end{equation}
\begin{equation}
\begin{split}
\Delta_{2}=&\frac{32 |\textbf{a}|^2 e^2}{(k.p_{f}) (k.p_{i})}\bigg[g_{V}^2 ((k_{f}.p_{f})(k.k_{i}) (k.p_{f})-(k_{f}.p_{i})(k.k_{i})(k.p_{f})+(k_{i}.p_{f})(k.k_{f})(k.p_{f}- k.p_{i})\\
&-(k_{f}.p_{f})(k.k_{i})(k.p_{i})+(k_{f}.p_{i})(k.k_{i})(k.p_{i})+2(k_{i}.k_{f})(k.p_{f})(k.p_{i})+(k_{i}.p_{i}) (k.k_{f})\\
&\times(-k.p_{f} + k.p_{i})-2(k.k_{f})(k.k_{i})m^2 +2(k.k_{f})(k.k_{i})(p_{i}.p_{f})-2(k.k_{i})(k.p_{i})(\eta_{1}.k_{f})\\
&\times(\eta_{1}.p_{f})- 2(k.k_{i})(k.p_{f})(\eta_{1}.k_{f})(\eta_{1}.p_{i})+4(k.k_{f}) (k.k_{i}) (\eta_{1}.p_{f})( \eta_{1}.p_{i}))+g_{A}^2 ((k_{f}.p_{f})\\
&\times(k.k_{i})(k.p_{f})-(k_{f}.p_{i})(k.k_{i})(k.p_{f})+(k_{i}.p_{f})(k.k_{f})(k.p_{f}-k.p_{i})-(k_{f}.p_{f})(k.k_{i})(k.p_{i}) \\
&+(k_{f}.p_{i})(k.k_{i})(k.p_{i})+2(k_{i}.k_{f})(k.p_{f})(k.p_{i})+(k_{i}.p_{i})(k.k_{f})(-kp_{f} + kp_{i})+2 (k.k_{f})\\
&\times (k.k_{i})m^2+2(k.k_{f})(k.k_{i})(p_{i}.p_{f})-2(k.k_{i})(k.p_{i}) (\eta_{1}.k_{f})(\eta_{1}.p_{f})-2(k.k_{i})(k.p_{f})\\
&\times(\eta_{1}.k_{f}) (\eta_{1}.p_{i})+4 (k.k_{f})(k.k_{i}) (\eta_{1}.p_{f}) (\eta_{1}.p_{i}))-2 g_{A} g_{V} ((k_{i}.p_{f})(k.k_{f})(k.p_{f} + k.p_{i})\\
& - (k_{i}.p_{i})(k.k_{f})(kp_{f} + kp_{i})+(k.k_{i})(-(k_{f}.p_{f})(k.p_{f} + k.p_{i}) + (k_{f}.p_{i})(k.p_{f} + k.p_{i})\\
&-2 (k.p_{i}) (\eta_{1}.k_{f})(\eta_{1}.p_{f})+2(k.p_{f})(\eta_{1}.k_{f})(\eta_{1}.p_{i}))))\bigg],
\end{split}
\end{equation}
\begin{equation}
\begin{split}
\Delta_{3}=&\frac{32 |\textbf{a}|^2 e^2}{(k.p_{f}) (k.p_{i})}\bigg[g_{V}^2 ((k_{f}.p_{f})(k.k_{i})(k.p_{f})-(k_{f}.p_{i})(k.k_{i})(k.p_{f})+(k_{i}.p_{f})(k.k_{f})\\
&\times(k.p_{f} - k.p_{i}) -(k_{f}.p_{f})(k.k_{i})(k.p_{i})+(k_{f}.p_{i})(k.k_{i})(k.p_{i})+2(k_{i}.k_{f})(k.p_{f})(k.p_{i})\\
&+(k_{i}.p_{i}) (k.k_{f})(-k.p_{f} + k.p_{i})-2(k.k_{f})(k.k_{i}) m^2 + 2(k.k_{f})(k.k_{i})(p_{i}.p_{f}) -2(k.k_{i})\\
&\times(k.p_{i})(\eta_{2}.k_{f})(\eta_{2}.p_{f})-2(k.k_{i})(k.p_{f})(\eta_{2}.k_{f})(\eta_{2}.p_{i})+4 (k.k_{f})(k.k_{i})(\eta_{2}.p_{f})(\eta_{2}.p_{i})\\
&+g_{A}^2 ((k_{f}.p_{f})(k.k_{i})(k.p_{f})-(k_{f}.p_{i})(k.k_{i})(k.p_{f})+(k_{i}.p_{f})(k.k_{f})(kp_{f} - kp_{i})-(k_{f}.p_{f})\\
&\times(k.k_{i})(k.p_{i}) +(k_{f}.p_{i})(k.k_{i})(k.p_{i}) + 2 (k_{i}.k_{f})(k.p_{f})(k.p_{i}) + (k_{i}.p_{i})(k.k_{f}) (-k.p_{f} + k.p_{i})\\
&+2(k.k_{f})(k.k_{i})m^2 +2(k.k_{f})(k.k_{i})(p_{i}.p_{f})-2(k.k_{i})(k.p_{i})((\eta_{2}.k_{f})((\eta_{2}.p_{f})-2(k.k_{i})(k.p_{f})\\
&\times((\eta_{2}.k_{f})((\eta_{2}.p_{i})+4(k.k_{f})(k.k_{i})((\eta_{2}.p_{f})((\eta_{2}.p_{i}))-2 g_{A} g_{V} ((k_{i}.p_{f})(k.k_{f})(k.p_{f} + k.p_{i})\\
&-(k_{i}.p_{i})(k.k_{f}) (k.p_{f} + k.p_{i})+(k.k_{i})(-(k_{f}.p_{f})(k.p_{f} + k.p_{i})+(k_{f}.p_{i}) (k.p_{f} + k.p_{i})\\
&-2 (k.p_{i})(\eta_{2}.k_{f})(\eta_{2}.p_{f})+2(k.p_{f})(\eta_{2}.k_{f})(\eta_{2}.p_{i})))\bigg],
\end{split}
\end{equation}
\begin{equation}
\begin{split}
\Delta_{4}=&\frac{-16 e^2}{(k.p_{f})^2 (k.p_{i})^2}\bigg[2|\textbf{a}|^2(k.k_{i})(k.p_{f})(k.p_{i})\Big(g_{A}^2 ((k.p_{i})(\eta_{1}.p_{f})(\eta_{2}.k_{f})+(k.p_{f})(\eta_{1}.p_{i})(\eta_{2}.k_{f})\\
&+(k.p_{i})(\eta_{1}.k_{f})(\eta_{2}.p_{f})-2 (k.k_{f})(\eta_{1}.p_{i})(\eta_{2}.p_{f})+ (k.p_{f})(\eta_{1}.k_{f})(\eta_{2}.p_{i})-2 (k.k_{f})(\eta_{1}.p_{f})\\
&\times(\eta_{2}.p_{i}))+g_{V}^2 ((k.p_{i})(\eta_{1}.p_{f})(\eta_{2}.k_{f})+(k.p_{f})(\eta_{1}.p_{i})(\eta_{2}.k_{f})+(k.p_{i})(\eta_{1}.k_{f})(\eta_{2}.p_{f})\\
&-2(k.k_{f})(\eta_{1}.p_{i})(\eta_{2}.p_{f})+(k.p_{f})(\eta_{1}.k_{f})(\eta_{2}.p_{i})-2(k.k_{f})(\eta_{1}.p_{f})(\eta_{2}.p_{i}))+2 g_{A} g_{V}\\
&\times (-(k.p_{i})((\eta_{1}.p_{f})(\eta_{2}.k_{f})+(\eta_{1}.k_{f})(\eta_{2}.p_{f}))+ (k.p_{f})((\eta_{1}.p_{i})(\eta_{2}.k_{f})+(\eta_{1}.k_{f})(\eta_{2}.p_{i})))\Big)\\
&-i \Big(g_{A} g_{V}|\textbf{a}|^2 \Big((k_{i}.p_{i})(k.p_{f}) ((k.p_{f})^2- 2(k.p_{f})(k.p_{i})+2(k.p_{i})^2)-(k.p_{f}- k.p_{i})((k_{i}.p_{f})\\
&\times(k.p_{f})(k.p_{i})+ (k.k_{i})(k.p_{f} + k.p_{i})(p_{i}.p_{f}))\Big)\epsilon(\eta_{1},\eta_{2},k,k_{f})+g_{A} g_{V} |\textbf{a}|^2 \Big((k_{f}.p_{i})(k.p_{f})((k.p_{f})^2\\
&-2 (k.p_{f})(k.p_{i})+2(k.p_{i})^2) -(k.p_{f} -k.p_{i})((k_{f}.p_{f})(k.p_{f})(k.p_{i})+(k.k_{f})(k.p_{f} + k.p_{i})\\
&\times(p_{i}.p_{f}))\Big)\epsilon(\eta_{1},\eta_{2},k,k_{i})-2|\textbf{a}|^2(g_{A}^2+g_{V}^2)(k.p_{f})(k.p_{i}) (k.p_{f}^2 + k.p_{i}^2)\epsilon(\eta_{1},\eta_{2},k_{f},k_{i})\\
&+2|\textbf{a}|^2(g_{A}^2 + g_{V}^2)(k.p_{f})(k.p_{i}) ((k.p_{f})(\eta_{2}.p_{f})+(k.p_{i})(\eta_{2}.p_{i}))\epsilon(\eta_{1},k,k_{f},k_{i})\\
&-2 |\textbf{a}|^2 (g_{A}^2 + g_{V}^2)(k.p_{f})(k.p_{i})((k.p_{f})(\eta_{1}.p_{f} )+(k.p_{i})(\eta_{1}.p_{i}))\epsilon(\eta_{2},k,k_{f},k_{i})-g_{A} g_{V} |\textbf{a}|^2(k.p_{i})\\
&\times(-((k_{i}.p_{i})(k.k_{f})+(k_{f}.p_{i})(k.k_{i}))(k.p_{i}) + 2(k_{i}.k_{f})(k.p_{f})(k.p_{f} + k.p_{i}))\\
&\times\epsilon(\eta1,\eta_{2},k,p_{f})+g_{A} g_{V} |\textbf{a}|^2 (k.k_{i})(k.p_{i}) (k.p_{f}^2 + (k.p_{f})(k.p_{i}) + k.p_{i}^2)\epsilon(\eta_{1},\eta_{2},k_{f},p_{f})\\
&-2 |\textbf{a}|^2 g_{A} g_{V} (k.k_{i})(k.p_{f})(k.p_{i})(\eta_{2}.p_{i})\epsilon(\eta_{1},k,k_{f},p_{f})-|\textbf{a}|^2 g_{A} g_{V} (k.k_{i})(k.p_{f}^2-2 (k.p_{f})(k.p_{i})\\
&-k.p_{i}^2)(\eta_{1}.p_{i})\epsilon(\eta_{2},k,k_{f},p_{f})+g_{A} g_{V} |\textbf{a}|^2(k.k_{f})(k.p_{i})(k.p_{f}^2 +(k.p_{f})(k.p_{i})+k.p_{i}^2)\\
&\times\epsilon(\eta_{1},\eta_{2},k_{i},p_{f})-|\textbf{a}|^2 g_{A} g_{V}(k.p_{f})(k.p_{i})((k.p_{f})(\eta_{2}.k_{f})-(k.p_{i})(\eta_{2}.k_{f})+2(k.k_{f})\\
&\times(\eta_{2}.p_{i}))\epsilon(\eta_{1},k,k_{i},p_{f})+|\textbf{a}|^2 g_{A} g_{V} ((k.p_{f})(k.p_{f} - k.p_{i})(k.p_{i})(\eta_{1}.k_{f})+(k.k_{f})\\
&\times(-k.p_{f}^2+2(k.p_{f})(k.p_{i})+ k.p_{i}^2)(\eta_{1}.p_{i}))\epsilon(\eta_{2},k,k_{i},p_{f})-g_{A} g_{V} |\textbf{a}|^2(k.p_{f})^2((k_{i}.p_{f})(k.k_{f})\\
&+(k_{f}.p_{f})(k.k_{i})+ 2(k_{i}.k_{f})(k.p_{f})-4(k_{i}.k_{f})(k.p_{i}))\epsilon(\eta_{1},\eta_{2},k,p_{i}) -2 g_{A} g_{V} |\textbf{a}|^2(k.k_{i})\\
&\times(k.p_{f})^2(k.p_{i})\epsilon(\eta_{1},\eta_{2},k_{f},p_{i})+|\textbf{a}|^2 g_{A} g_{V} (k.k_{i}) (k.p_{f}^2 + 2(k.p_{f})(k.p_{i})-k.p_{i}^2)\\
&\times(\eta_{2}.p_{f})\epsilon(\eta_{1},k,k_{f},p_{i})-2 |\textbf{a}|^2 g_{A} g_{V}(k.k_{i})(k.p_{f})(k.p_{i}) (\eta_{1}.p_{f})\epsilon(\eta_{2},k,k_{f},p_{i})-2 g_{A} g_{V} |\textbf{a}|^2\\
&\times(k.k_{f})(k.p_{f})^2(k.p_{i})\epsilon(\eta_{1},\eta_{2},k_{i},p_{i})-|\textbf{a}|^2 g_{A} g_{V} ((k.p_{f})^3(\eta_{2}.k_{f})+(k.k_{f})(-k.p_{f}^2 -2 (k.p_{f})\\
&\times(k.p_{i})+k.p_{i}^2)(\eta_{2}.p_{f}))\epsilon(\eta_{1},k,k_{i},p_{i})+|\textbf{a}|^2 g_{A} g_{V}(k.p_{f}) ((k.p_{f})^2(\eta_{1}.k_{f})-2(k.k_{f})\\
&\times(k.p_{i})(\eta_{1}.p_{f}))\epsilon(\eta_{2},k,k_{i},p_{i})+|\textbf{a}|^2 g_{A} g_{V}(k.k_{i})(k.p_{i}^2) (\eta_{2}.k_{f})\epsilon(\eta_{1},k,p_{f},p_{i}) \\
&-|\textbf{a}|^2 g_{A} g_{V}(k.k_{i})(k.p_{f})^2(\eta_{1}.k_{f})\epsilon(\eta_{2},k,p_{f},p_{i})\Big)\bigg].
\end{split}
\end{equation}
To compute analytically the terms $\epsilon(a,b,c,d)$ that appears during the evaluation of the two traces mentioned in eq. (\ref{IDCS}), we use the Grozin convention:
\begin{equation}
\epsilon_{0123}=1,
\end{equation}
which implies that $\epsilon_{\rho\sigma\mu\nu} = -1\, (1)$ for an odd (even) permutation of the Lorentz indices and $\epsilon_{\rho\sigma\mu\nu}=0$  otherwise. For more information, we give here how to compute one of the tensors that appear in the coefficient $\Delta_{4}$:
\begin{equation}
\begin{split}
\epsilon(\eta_{1},\eta_{2},k,k_{f})=& \epsilon_{\rho\sigma \mu\nu} \eta_{1}^{\rho}\eta_{2}^{\sigma}k^{\mu}k_{f}^{\nu}\\
=& \epsilon_{1203}~ \eta_{1}^{1}\eta_{2}^{2}k^{0}k_{f}^{3}+\epsilon_{1230}~ \eta_{1}^{1}\eta_{2}^{2}k^{3}k_{f}^{0}\\
=& \eta_{1}^{1}\eta_{2}^{2}(k^{0}k_{f}^{3}-k^{3}k_{f}^{0})\\
=&\omega(|\textbf{q}_{i}|\cos(\theta_{i})-|\textbf{q}_{f}|\cos(\theta_{f})-E_{i}+n \omega-E_{f}).
\end{split}
\end{equation}


\begin{thebibliography}{99}
\bibitem{highlaser} Bahk S W, Rousseau P, Planchon T A, Chvykov V, Kalintchenko G, Maksimchuk A, Mourou G A and Yanovsky V (2004) Generation and characterization of the highest laser intensities ($10^{22}~\text{W/cm}^{2}$) \emph{Opt. Lett.} \textbf{29} 2837.
\bibitem{laser} H. Kiriyama \textit{et al.}, High-contrast high-intensity repetitive petawatt laser, Opt. Lett. \textbf{43}, 2595 (2018).
\bibitem{laser1} C. N. Danson \textit{et al.}, Petawatt and exawatt class lasers worldwide, High Power Laser Sci. Eng. \textbf{7}, e54 (2019).
\bibitem{laser3} G.A. Mourou, T.Tajima, and S. V. Bulanov, Optics in the relativistic regime, \emph{ Rev. Mod.Phys.} \textbf{78}, 309 (2006);http://dx.doi.org/10.1103/RevModPhys.78.309.
\bibitem{test1} Kumita T \textit{et al} (2006) Observation of the nonlinear effect in relativistic Thomson scattering of electron and laser beams \emph{Laser Phys.} \textbf{16} 267.
\bibitem{test2} Burke D L \textit{et al} (1997) Positron production in multiphoton light-by-light scattering \emph{Phys. Rev. Lett.} \textbf{79} 1626.
\bibitem{test3} Bula C \textit{et al} (1996) Observation of nonlinear effects in Compton scattering \emph{Phys. Rev. Lett.} \textbf{76} 3116.
\bibitem{faisal} Faisal F H M (1987) \textit{Theory of Multiphoton Processes} (New York: Plenum).
\bibitem{mittleman} Mittleman M H (1993) \textit{Introduction to the Theory of Laser-Atom Interactions} (New York: Plenum).
\bibitem{fedorov} Fedorov M V (1997) \textit{Atomic and Free Electrons in a Strong Light Field} (Singapore: World Scientific).
\bibitem{szymanowski1} Szymanowski C and Maquet A (1998) Relativistic signatures in laser-assisted scattering at high field intensities \emph{Opt. Express} \textbf{2} 262.
\bibitem{szymanowski2} Szymanowski C, V\'eniard V, Ta\"{i}eb R, Maquet A and Keitel C H (1997) Mott scattering in strong laser fields \emph{Phys. Rev. A} \textbf{56} 3846.
\bibitem{manaut2003} Attaourti Yand and Manaut B, (2003) comment on Mott scattering in strong laser fields  \emph{Phys. Rev. A} \textbf{68} 067401.
\bibitem{li2003} Li S M, Berakdar J, Chen J and  Zhou Z F (2003) Mott scattering in the presence of a linearly polarized laser field \emph{Phys. Rev. A} \textbf{67} 063409.
\bibitem{attaourti2004} Attaourti Y,  Manaut B and Taj S (2004) Mott scattering in an elliptically polarized laser field \emph{Phys. Rev. A} \textbf{70} 023404.
\bibitem{dahiri} Dahiri I, Jakha M, Mouslih S, Manaut B, Taj S and Attaourti Y (2021) Elastic electron-proton scattering in the presence of a circularly polarized laser field \emph{Laser Phys. Lett.} \textbf{18} 096001.
\bibitem{liu2014} Liu A H and Li S M (2014) Relativistic electron scattering from a freely movable proton in a strong laser field \emph{Phys. Rev. A} \textbf{90} 055403.
\bibitem{wang2019} Wang N, Jiao L and Liu A (2019) Relativistic electron scattering from freely movable proton/$\mu^{+}$ in the presence of strong laser field \emph{Chin. Phys. B} \textbf{28} 093402.
\bibitem{du2018} Du W Y, Zhang P F and Wang B H (2018) New phenomena in laser-assisted scattering of an electron by a muon \emph{Front. Phys.} \textbf{13} 133401.
\bibitem{Nonresonant} E.A. Padusenko, S.P. Roshchupkin, and A.I. Voroshilo (2009) Nonresonant scattering of relativistic electron by
relativistic muon in the pulsed light field \emph{Laser Phys. Lett} \textbf{6} 3, 242–251.
\bibitem{Yahya} Y. Mekaoui, M. Jakha, S. Mouslih, B. Manaut, R. Benbrik, S. Taj (2021) Relativistic elastic scattering of an electron by a muon in the field of a circularly polarized electromagnetic wave 	arXiv:2110.06695.
\bibitem{Weinberg} Weinberg, S (1967) A Model of Leptons \emph{Phys. Rev. Lett} \textbf{19} 21, 1264–66.
\bibitem{Salam} Salam A, Ward J. C. (1959) Weak and electromagnetic interactions \emph{Nuovo Cimento}  \textbf{11} 4,  568–577.
\bibitem{jakha} Jakha M, Mouslih S, Taj S and Manaut B (2021) Laser effect on the final products of $Z$-boson decay \emph{Laser Phys. Lett.} \textbf{18} 016002.
\bibitem{hadwdecay} Jakha M, Mouslih S, Taj S, Attaourti Y and Manaut B (2021) Influence of intense laser fields on measurable quantities in $W^-$-boson decay \textit{Chin. J. Phys.} \url{https://doi.org/10.1016/j.cjph.2021.09.011}.
\bibitem{ouhammou} Ouhammou M, Ouali M, Taj S and Manaut B (2021) Higgs-strahlung boson production in the presence of a circularly polarized laser field \emph{Laser Phys. Lett.} \textbf{18} 076002.
\bibitem{Ouali} Ouali M, Ouhammou M, Mekaoui Y, Taj S and Manaut B (2021) \textit{Chinese Journal of Physics}, https://doi.org/10.1016/j.cjph.2021.10.007. 
\bibitem{Muller} Muller, Z. Hatsagortsyan, C.H. Keitel, (2006), \emph{Phys. Rev. D}, \textit{74} 074017, http://dx.doi.org/10.1103/.
\bibitem{vilain} P. Vilain \textit{et al.} (CHARM II Collaboration), Precision measurement of electroweak parameters from the scattering of muon-neutrinos on electrons, \emph{Phys. Lett. B} \textbf{335}, 246 (1994).
\bibitem{hasert} F. J. Hasert \textit{et al.} (Gargamelle Collaboration), Search for elastic muon-neutrino electron scattering, Phys. Lett. B \textbf{46}, 121 (1973).
\bibitem{tomalak} O. Tomalak and R. J. Hill, Theory of elastic neutrino-electron scattering, \textit{Phys. Rev. D} \textbf{101}, 033006 (2020).
\bibitem{marciano} W. J. Marciano and Z. Parsa, Neutrino-electron scattering theory, J. Phys. G: Nucl. Part. Phys. \textbf{29}, 2629 (2003).
\bibitem{bai2012} L. Bai, M. Y. Zheng, and B. H. Wang, Multiphoton processes in laser-assisted scattering of a muon neutrino by an electron,  \emph{Phys. Rev. A} \textbf{85}, 013402 (2012).
\bibitem{ElAsri} S. El Asri, S. Mouslih,  M. Jakha,  B. Manaut, Y. Attaourti, S. Taj and R. Benbrik, Elastic scattering of a muon neutrino by an electron in the presence of a circularly polarized laser field, 2021 \emph{Phys. Rev. D}, \textit{104}, 113001.
\bibitem{greiner} W. Greiner and B. M\"{u}ller, \textit{Gauge Theory of Weak Interactions}, 4th ed. (Springer, Berlin, 2009).
\bibitem{MohaPhLettB} M. Ouali, M. Ouhammou, S. Taj, B. Manaut , R. Benbrik  Laser-assisted charged Higgs pair production in inert Higgs Doublet Model (IHDM, 2021, \emph{Physics Letters B}, \textbf{823}, 136761.
\bibitem{pdg2020} P. A. Zyla \textit{et al.} (Particle Data Group), Review of Particle Physics, Prog. Theor. Exp. Phys. \textbf{2020}, 083C01 (2020).
\bibitem{PDGroup2020} Zyla P A et al (Particle Data Group) 2020 \emph{Prog. Theor. Exp. Phys}, \textit{2020} 083C01.
\bibitem{volkov} D. M. Volkov, On a class of solutions of the Dirac equation, Z. Phys. \textbf{94}, 250 (1935).
\bibitem{feyncalc1} R. Mertig, M. B\"{o}hm, and A. Denner, Feyn Calc - Computer-algebraic calculation of Feynman amplitudes, Comput. Phys. Commun. \textbf{64}, 345 (1991).
\bibitem{feyncalc2} V. Shtabovenko, R. Mertig, and F. Orellana,  New developments in FeynCalc 9.0,  Comput. Phys. Commun. \textbf{207}, 432 (2016). 
\bibitem{feyncalc3} V. Shtabovenko, R. Mertig, and F. Orellana, FeynCalc 9.3: New features and improvements, Comput. Phys. Commun. \textbf{256}, 107478 (2020).
\bibitem{Watson} N. M. Kroll and K. M. Watson, Charged-particle scattering in the presence of a strong electromagnetic wave, Phys. Rev. A \textbf{8}, 804 (1973).
\end{thebibliography}
\end{document}